%% file: Thesis.tex
\documentclass[12pt,a4paper,bold]{thesis}

\usepackage{amsmath}
\usepackage{amsthm}
\usepackage{amssymb}
\usepackage{setspace}
\usepackage{apacite}
\usepackage[authoryear]{natbib}

\usepackage[colorlinks=true,allcolors=blue]{hyperref}
\usepackage[utf8]{inputenc}
\usepackage{mathtools}
\allowdisplaybreaks
\usepackage{graphicx}
\graphicspath{ {figures/} }
\usepackage{array}
\usepackage{tabularx}
\usepackage{subcaption}
\usepackage{float}
\usepackage{enumitem}
\usepackage[bottom=1in]{geometry}

\usepackage{tikz}
\usepackage{scalerel}
\usepackage{pict2e}
\usepackage{tkz-euclide}
\usetikzlibrary{calc}
\usetikzlibrary{patterns,arrows.meta}
\usetikzlibrary{shadows}
\usetikzlibrary{external}
\usepackage{pgfplots}
\pgfplotsset{compat=newest}
\usepgfplotslibrary{statistics}
\usepgfplotslibrary{fillbetween}
\usepackage{xcolor}
\theoremstyle{thm}

\theoremstyle{definition}

\usepackage{graphicx}
\usepackage{epstopdf}

\usepackage[toc, page]{appendix}

\providecommand{\appendixname}{Appendices}

\newcommand{\head}[1]{\newpage
\vspace{3em}
\begin{center}
\LARGE{\MakeUppercase{\textbf{#1}}}
\end{center}
\vspace{3em}
\addcontentsline{toc}{chapter}{#1}
}
\renewcommand{\listfigurename}{LIST OF FIGURES}
\newcommand{\thesistitle}{Pulse Signal Simulation of Pulsars}
\newcommand{\studentname}{Jalormi Brahmachari}
\newcommand{\studentrollno}{21127}
\newcommand{\advisorname}{Dr Mayuresh Surnis}

\newcommand{\subject}{Physics}
\newcommand{\department}{Physics}
\newcommand{\thesisdate}{April 2026}

\def\maketitle{
\begin{titlepage}
\begin{center}
\begin{doublespace}
\textbf{\MakeUppercase{\LARGE{\thesistitle}}} \\~\\~\\
\normalsize{\textbf{A THESIS}} \\
\normalsize{\textit{submitted in partial fulfillment of the requirements}} \\
\normalsize{\textit{for the award of the dual degree of}} \\~\\
\large{\textbf{Bachelor of Science - Master of Science}} \\
\normalsize{\textit{in}} \\
\large{\textbf{\MakeUppercase{\subject}}} \\
\normalsize{\textit{by}} \\
\large{\textbf{\MakeUppercase{\studentname}}} \\
\normalsize{\textbf{(\studentrollno)}} \\~\\~\\
\end{doublespace}

\centerline{\includegraphics[scale=0.20]{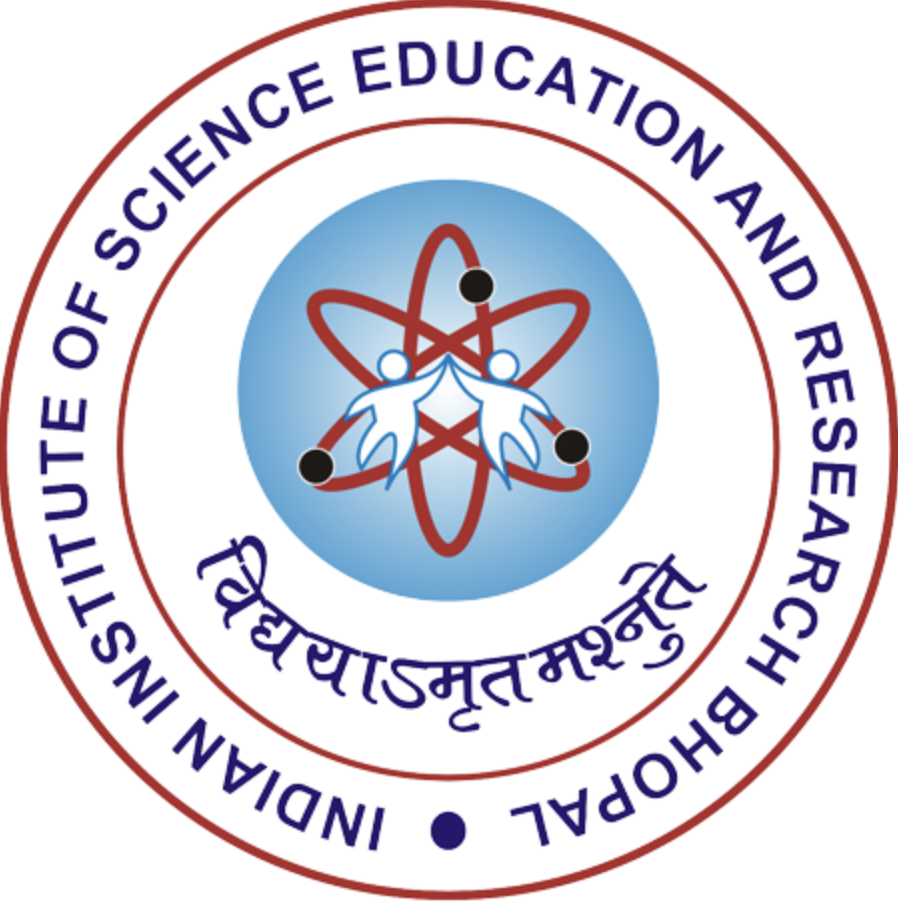}}
\textbf{DEPARTMENT OF \MakeUppercase{\department} \\ 
INDIAN INSTITUTE OF SCIENCE EDUCATION AND RESEARCH BHOPAL\\ %Flows onto two lines
BHOPAL - 462066} \\~\\
\textbf{\thesisdate}
\end{center}
\end{titlepage}
}

\begin{document}
\maketitle

\pagenumbering{roman}

% ------------------------------
\head{Certificate}

\begin{figure}
    \centering
    \includegraphics{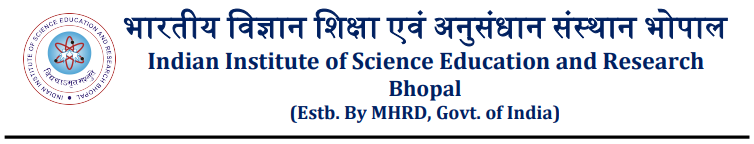}
\end{figure}

\thispagestyle{empty}

This is to certify that {\bf \studentname}, BS-MS (Dual Degreee) student in Department of {\subject}, has completed bonafide work on the thesis entitled {\bf `\thesistitle'} under my supervision and guidance.

\vspace{10em}

\textbf{\thesisdate \hfill \advisorname \\ IISER Bhopal \hfill (IISER Bhopal)}

% In case of no co-supervisor: Remove the name, institute and corresponding "hfill".

\vfill

\begin{center}
\begin{tabular}{ccc}
\textbf{Committee Member} & \textbf{Signature} & \textbf{Date} \\
\\
Member 1 & \rule{12em}{0.4pt} & \rule{6em}{0.4pt} \\
\\
Member 2 & \rule{12em}{0.4pt} & \rule{6em}{0.4pt} \\
\\
Member 3 & \rule{12em}{0.4pt} & \rule{6em}{0.4pt} \\
\end{tabular}
\end{center}

% ------------------------------
\head{Academic Integrity and Copyright Disclaimer}

I hereby declare that this project report is my own work and due
acknowledgement has been made wherever the work described is based on the
findings of other investigators. This report has not been accepted for the award of any other degree or diploma at IISER Bhopal or any other educational institution. I also declare that I have adhered to all principles of academic honesty and integrity and have not misrepresented or fabricated or falsified any idea/data/fact/source in my submission.\\

I certify that all copyrighted material incorporated into this document is in
compliance with the Indian Copyright (Amendment) Act (2012) and that I have
received written permission from the copyright owners for my use of their work,
which is beyond the scope of the law. I agree to indemnify and safeguard IISER
Bhopal from any claims that may arise from any copyright violation.

\vfill

\textbf{\thesisdate \hfill \studentname \\ IISER Bhopal}

% ------------------------------
\head{Acknowledgement}
Firstly, I would like to express my gratitude towards my thesis supervisor, \textbf{Dr. Mayuresh Surnis}, for his constant support and motivation throughout this entire duration. His wonderful ideas for this thesis and constant words of wisdom have helped me complete this work successfully. He has been my mentor since I was in the second year and I would not have been so inclined towards Astrophysics if it were not for him. He introduced me to this beautiful realm of physics which continues to amaze me and keeps me inspired to pursue a career in no other field. 

I want to thank my parents for their immense love and for sticking by me in the times I lost my self-confidence and was afraid to follow my dreams. This would not have been possible without their belief in me and their constant advice which has kept me on the right path of life. I would also like to thank all my grandparents, especially my Nana, my Kaka, and my cousin, Kshitish Bhai for always supporting and motivating me. 

I would also like to thank the PhDs in our lab Hemanga Bhaiya, Kunjal didi, Kaustubh, Priyanshi and especially \textit{Abhinandan Bhaiya}, for being a constant source of support and guidance. I am grateful for the cheerful environment that is always present in the lab because of these people who made working fun. 
Without them, it would have been a difficult task to achieve this.  

Lastly, I would like to thank all my dear friends whom I met in this institute. The family that I made here, Maitree, Akshi, and Harshal, the people who have been my constant all these years and have always been there for me, no matter what. I would like to thank my roommate, Gargi for all the amazing time we shared. As much as I would like to name all the wonderful people I have met here, I would mention a few names, Omkar, Ishika, Ishan Thoke, Hemant, Avantika, Niharika, Mazia, Revati, Saurav. Special thanks to all my society friends, as well who have been with me since my childhood.  

\vspace{7em}

\begin{flushright}
    {\bf \studentname}
\end{flushright}

% ------------------------------
\head{Abstract}
The present thesis is a work in progress on improving the simulation of integrated pulse profiles of pulsars. Pulsars, highly magnetized rotating neutron stars, serve as precise cosmic clocks useful for studying gravity and the interstellar medium. Although periodic pulses from many pulsars are observed and modeled at different radio frequencies, a robust realistic simulation of their profiles utilizing their parameters remain underdeveloped. The present study focuses on developing a physically consistent model to reproduce observed pulse shapes across
multiple frequencies. This study will be useful in coherently understanding the various aspects of the pulsar emission mechanism. These integrated pulse profiles are the time-averaged properties of pulsar emission and are obtained after averaging individual pulses over a few thousand rotations. Observed to be highly stable, these are indicative of the global properties of the pulsar magnetosphere. By refining simulation techniques and incorporating the propagation effects of the interstellar medium, this work will prove to be helpful in generating synthetic, yet realistic models. The results are expected to contribute to the interpretation of observational data and to the development of improved timing and emission models for pulsars.

% ------------------------------
\head{List of Symbols or Abbreviations}

\begin{center}
\begin{tabular}{l@{\hspace{7em}}l@{}} \smallskip
	\textbf{PSR} & Pulsating Radio Source  \\ \smallskip
	\textbf{MSP} & Millisecond Pulsar  \\ \smallskip
    \textbf{GW} & Gravitational Waves \\ \smallskip
    \textbf{SNR} & Supernova Remnant \\ \smallskip
    \textbf{S/N} & Signal to Noise Ratio \\ \smallskip
    \textbf{TOA} & Time of Arrival \\ \smallskip
    \textbf{PsrSigSim} & Pulsar Signal Simulation  \\ \smallskip
    \textbf{ISM} & Interstellar Medium   \\ \smallskip
\end{tabular}
\end{center}

% ------------------------------

\listoffigures
\addcontentsline{toc}{chapter}{List of Figures}

% \listofmyequations

% ------------------------------
% \head{List of Tables}
% \begin{center}
% \begin{tabular}{l@{\hspace{7em}}r@{}} \smallskip
% 	Nonlinear Model Results & 5 \\ \smallskip
% \end{tabular}
% \end{center}
\tableofcontents

% ------------------------------

\chapter{INTRODUCTION} \pagenumbering{arabic}
\label{ch:introduction} 
\input{Chapters/chap1}

\chapter{Canonical Pulsars and Millisecond Pulsars} \label{ch:ch2}

\input{Chapters/chap2}

\chapter{Observational data} \label{ch:ch3}

\input{Chapters/chap3}
\newpage

\chapter{Pulse Profile Simulator} \label{ch:ch4}

\input{Chapters/chap4}
\newpage

\chapter{Results and Discussion} \label{ch:ch5}

\input{Chapters/chap5}

% -----------------------------
\begin{appendices}
\renewcommand{\thesection}{\Roman{section}}
\section{\texttt{PSRCHIVE}}
\textbf{\texttt{PSRCHIVE}}\footnote{https://psrchive.sourceforge.net} is an open source C++ development library for the analysis of pulsar astronomical data. It implements an extensive range of algorithms for use in pulsar timing, polarimetric calibration, RFI mitigation, etc. These tools are utilized by a powerful suite of user-end programs that come with the library. The software is described in detail by \cite{Hotan2004PSRFITS}. These deal transparently and simultaneously with multiple data storage formats and enhances data portability and facilitates the adoption of the \texttt{PSRFITS} file format, a standard data storage format for pulsars. 

\texttt{PSRCHIVE} and \texttt{PSRFITS} were designed to form an object-oriented framework into which existing algorithms and data structure could be transplanted.  different telescopes and instruments require the storage of different types of information, including configuration parameters, observatory and instrumental status information, and other site-specific data. There is no way of knowing exactly what future systems might include, both \texttt{PSRCHIVE} and \texttt{PSRFITS} were implemented a by having generalised scheme for incorporating arbitrarily complex data extensions.

\begin{figure}[H]
    \centering
    \includegraphics[width=0.9\linewidth]{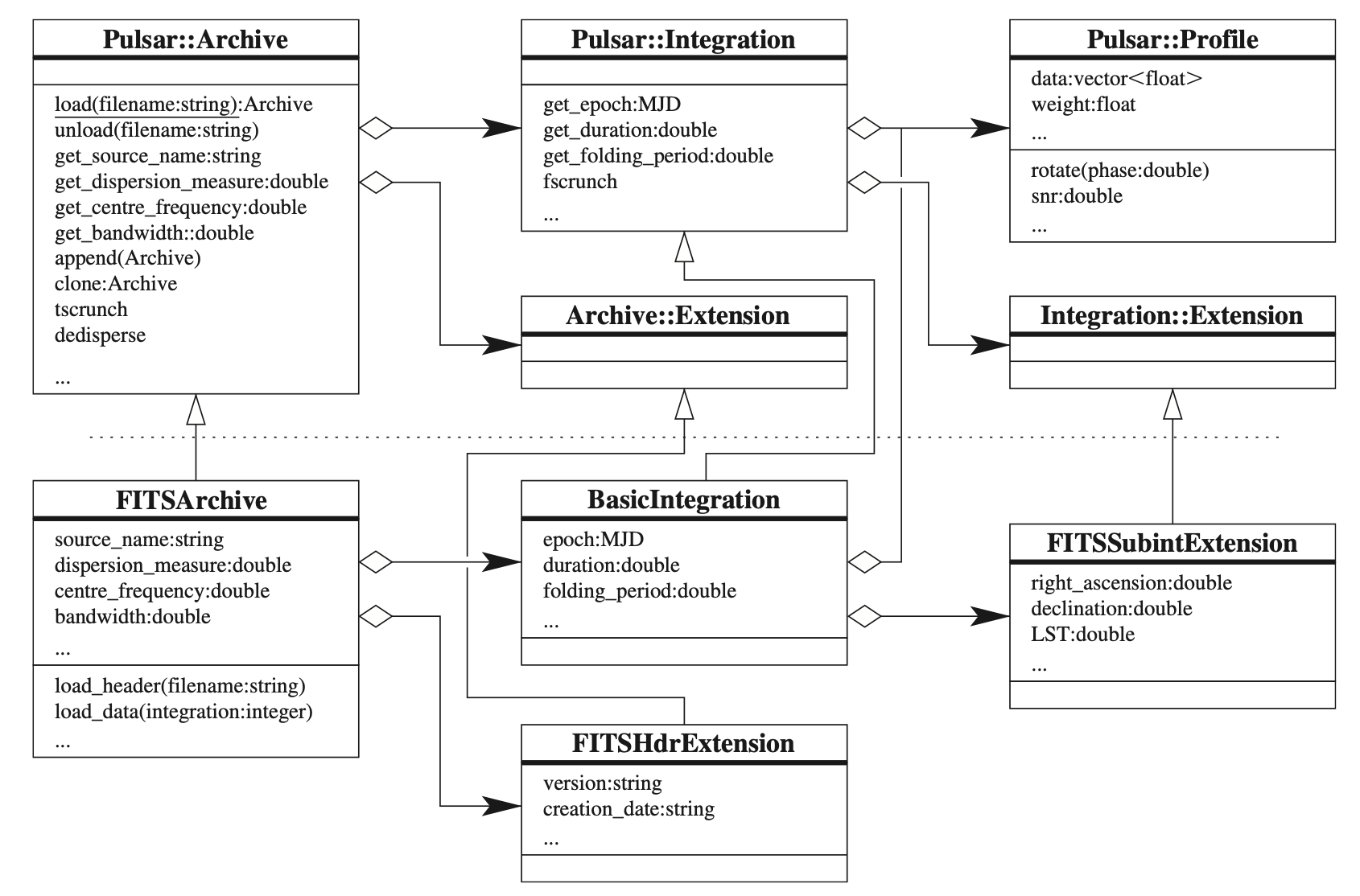}
    \caption{Class diagram of a portion of the \texttt{PSRCHIVE} library. The storage and access methods, as well as methods loading an dunloading data are described. The combined use of composition and inheritance enables complex structures and behaviours to be constructed using modular components \cite{Hotan2004PSRFITS}.}
    \label{fig:placeholder}
\end{figure}

\texttt{PSRCHIVE} provides with different applications, a few of them have been mentioned in the figure below, with their brief description that were included at the time of publication of \cite{Hotan2004PSRFITS}.

\begin{figure}[H]
    \centering
    \includegraphics[width=0.8\linewidth]{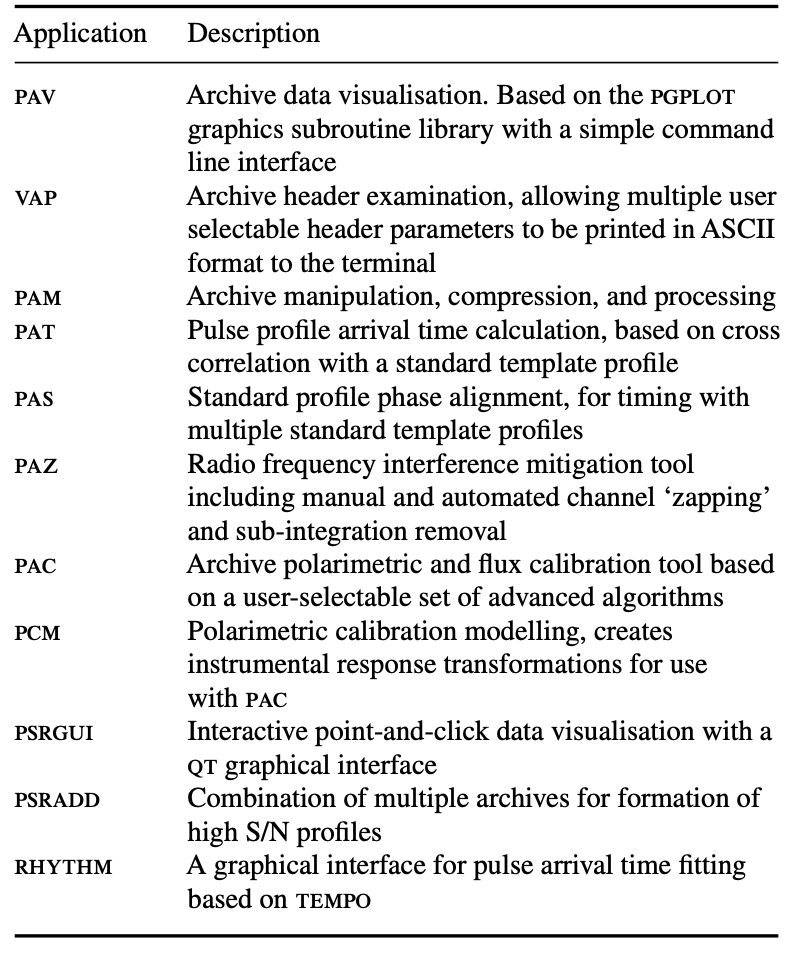}
    \caption{Standard Applications included with \texttt{PSRCHIVE}}
    \label{fig:placeholder}
\end{figure}

The combination of \texttt{PSRCHIVE} and \texttt{PSRFITS} provided a powerful, ready-to-use archive and reduction system for pulsar data, which was efficiently used in this thesis. 

\end{appendices}

% -----------------------------
% \nocite{*}
\bibliographystyle{apacite}
\bibliography{references}
\addcontentsline{toc}{chapter}{Bibliography}
\end{document}

%% file: Chapters/chap1.tex
\section{Neutron Stars} 
Neutron star is formed when a massive star runs out of fuel and collapses. This results in a supernova explosion of the massive star combined with gravitational collapse that compresses the core. The very central region of the star-- core--collapses crushing together every proton and electron into a neutron. 'If the core of the collapsing star is between about 1 and 3 solar masses, these newly created neutrons can stop the collapse and a neutron star is formed. Stars with higher masses continue to collapse into stellar-mass black hole.' 

A neutron star is one of the densest observable astrophysical objects with masses of 1 to 2 solar masses compressed to approximately 20 km in diameter. Since neutron stars began their existence as stars, they are found scattered throughout the galaxy and as stars they can be found by themselves or in a binary system with a companion. 

Neutron stars spin down over time, gradually losing their extreme rotational energy, primarily due to magnetic braking (magnetic dipole radiation and particle winds) and the emission of electromagnetic radiation. They are born spinning hundreds of times per second owing to the conservation of angular momentum during the collapse of the massive star's core but eventually slow down as their powerful magnetic fields interact with the surrounding environment, reducing their rotation rate. 

\section{Classification of Neutron Stars}
There are distinct observational classes of neutron stars. The emission spans the electromagnetic spectrum, and the radiative properties span a huge fraction of conceivable phase space. The great diversity of these neutron stars is based on the varied observational manifestations with their classification being recognized and summarized by \cite{kaspi2010} until now as : 

\begin{figure}[h!]
    \centering
    \includegraphics[width=0.5\linewidth]{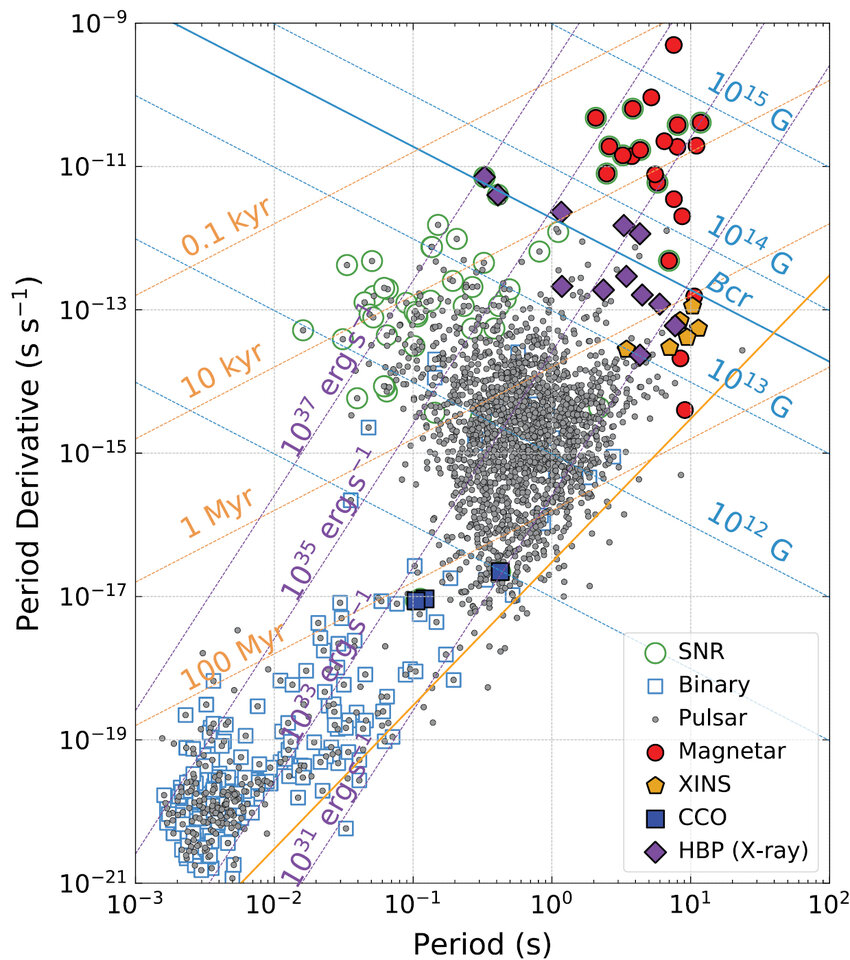}
    \caption[P-$\dot{P}$ diagram for distribution of known neutron stars]{\footnotesize P-$\dot{P}$ (spin period vs. period derivative) diagram for distribution of known neutron stars. Radio pulsars are shown as grey dots, magnetars as red circles, high-magnetic-field pulsars (HBP) as purple diamonds, X-ray isolated neutron stars (XINS) as orange pentagons, and compact central objects (CCOs) as blue squares. 
    Sources associated with supernova remnants (SNRs) are indicated by green circles, while binary systems are marked with blue outlines. 
    Lines of constant characteristic age (orange), surface magnetic field strength (blue), and spin-down luminosity (purple) are overlaid for reference. 
    The critical magnetic field is also indicated.}
    \label{fig:placeholder}
\end{figure}

\begin{enumerate}
    \item \textbf{Radio Pulsars (PSRs)} : Young rotation-powered neutron stars that emit regular pulses, primarily observed in the radio band, with periods ranging from tens of milliseconds to seconds.
    \item \textbf{Millisecond Pulsars (MSPs)} : Old neutron stars that have been spun up through accretion in binary systems, resulting in very short periods (1–30 ms) and weak magnetic fields.
    \item \textbf{Rotating Radio Transients (RRATs)} : Sources that emit sporadic, short-duration radio bursts with an underlying periodicity.
    \item \textbf{Isolated Neutron Stars (INSs)} : They have quasi-thermal x-ray emission with relatively low x-ray luminosity, great proximity, lack of radio counterpart, and relatively long periodicities as defining properties. 
    \item \textbf{Magnetars} : Neutron stars with extremely strong magnetic fields ($\sim 10^{14}–10^{15}$ G), exhibiting high-energy bursts in X-rays and gamma rays.
    \item \textbf{Compact Central Objects (CCOs)} : These are bright x-ray objects, a heterogeneous class. Named because of their central location in supernova remnants (SNRs). 
\end{enumerate}

\section{History of discovery of Pulsars}
Pulsars were first identified in 1967 at Cambridge University by graduate student Jocelyn Bell Burnell working with her supervisor Antony Hewish. While examining data from a newly built radio telescope, Bell noticed an unusual feature in the recordings, which she informally called a “bit of scruff.” At first, the regularity of the signal raised the possibility that it might originate from an artificial source, even prompting speculation about extraterrestrial communication. However, further investigation showed that the signal was natural in origin. Their discovery of the first pulsar was formally reported in February 1968.

Bell’s research project, supervised by Hewish, focused on studying quasars using the technique of interplanetary scintillation. Since quasars exhibit stronger scintillation than most radio sources, Hewish proposed this method as an effective way to investigate them and designed a dedicated radio telescope for this purpose.

Not long after the operations began, Bell detected an unusual pattern that did not resemble known scintillating sources or human-made noise. The signal appeared repeatedly from the same position in the sky and showed remarkable regularity. Bell and Hewish systematically eliminated possible sources such as radar echoes, satellites, television transmissions, and local structures near the telescope, but none could account for the phenomenon.

The emission consisted of narrow pulses separated by about 1.3 seconds, far too rapid to be associated with ordinary stars. As a joke, the team temporarily labeled the source “LGM-1” for “Little Green Men.” The extraterrestrial explanation was soon dismissed when Bell discovered additional pulsed sources from different regions of the sky, each with similar periodic behavior. By the end of 1967, four such objects had been identified.

In January 1968, Bell and Hewish submitted their results to Nature, and shortly before publication Hewish presented the findings in a seminar at Cambridge, even though the physical nature of the sources was still unknown. Within a year, many more pulsars had been found. Soon afterward, Thomas Gold proposed that pulsars are rapidly rotating neutron stars. Although neutron stars had been theoretically predicted in 1933, they were only confirmed observationally through the discovery of pulsars (\href{https://www.aps.org/archives/publications/apsnews/200602/history.cfm}{American Physical Society}).

\section{Pulsars as Tools for Probing Fundamental Physics}
Pulsars provide a wealth of information about neutron star physics, general relativity, the Galactic gravitational potential and magnetic field, the interstellar medium, celestial mechanics, planetary physics and even cosmology. 

Neutron stars are accessible to observation as pulsars and thus provide our only means of probing the most extreme states of matter in the present-day Universe, which in turn will enable a vast range of transforming science goals to be addressed, as mentioned in \cite{cordes2004} are :
\begin{itemize}
    \item Strong-field tests of gravity and the no-hair theorem of black holes (BHs). 
    \item Detection of a cosmological gravitational wave background.
    \item Mapping the complete structure of the Milky Way and revealing properties of the Galactic Center. 
    \item Probing the intergalactic medium in new ways. 
    \item Identifying the equation of state of superdense matter.
    \item Quantifying the role of magnetic fields and turbulence in core-collapse physics.
    \item Understanding the superfluid interiors and relativistic magnetospheres of neutron stars.
    \item Unraveling the evolutionary and dynamical histories and properties of all Galactic globular clusters.
    \item Discoveries of extra-solar planets. 
\end{itemize}

Given their remarkable rotational stability and diverse emission properties, pulsars--particularly millisecond pulsars--serve as powerful tools for precision astrophysics. In order to utilize these objects effectively, high-quality observational data and robust analysis techniques are essential. 

\section{Motivation and Objective}
The Python repository, \textbf{PsrSigSim}, developed by the NANOGrav collaboration, \footnote{https://psrsigsim.readthedocs.io/en/latest/readme.html}provides a flexible framework for simulating realistic pulsar signals. However, it has several limitations when modeling the frequency evolution of pulse profiles, particularly for known pulsars with complex and frequency-dependent structures.

Historically, it has been observed that the profiles of pulsars change with frequency over a broadband range. Their component separation, width, and amplitude change from low frequency to high frequency observations. This has been confirmed through multiple studies and different theoretical models have been proposed. This motivates the need to model and simulate these effects to understand the best possible mechanism for the observed evolution. Normal pulsars are known to follow specific trends because of their regular emission, whereas for millisecond pulsars, the typical profile evolution has been hardly observed and shows very little evolution of pulse width and component separation with frequency \citep{Kramer1999MSP}.   

In this work, rather than attempting a fully self-consistent modeling of frequency evolution--which remains challenging, particularly for millisecond pulsars--we adopt a data-driven approach. Integrated pulse profiles are decomposed into Gaussian components, and the resulting parameters are used to construct a library of profile descriptors reference at a standard frequency of 1 GHz. To the best of our knowledge, such a systematic library of Gaussian component parameters derived directly from observational data is not currently available in existing simulation frameworks, including PsrSigSim. This represents a key contribution of the present work. 

The resulting parameter library provides an empirical foundation for generating realistic pulse profiles within simulation environments. By anchoring simulations to observed profile characteristics, this approach enables the reproduction of source-specific pulse morphologies, while still allowing controlled exploration of propagation and instrumental effects. 

Such modeling offers insight into emission geometry, including emission altitudes and magnetospheric plasma properties, and their relation to observable quantities. The pulse profile plays a central role in determining the structure of the emission beam and the location of radio emission within the pulsar magnetosphere \citep{Mitra2016_MSPES, Rankin1993_VI}.

Furthermore, the parameter library enables the generation of realistic synthetic datasets for testing pulsar timing pipelines. These simulations facilitate the identification and quantification of time-of-arrival (TOA) uncertainties arising from instrumental and propagation effects. In cases where observational data are limited or incomplete, such synthetic profiles provide a robust alternative to validate analysis techniques.

Overall, this work establishes a framework that bridges observational data and simulation by providing empirically grounded pulse profile models, thereby contributing to improved interpretation of pulsar observations and the development of more realistic timing and emission studies.

This work is divided into four other chapters. Chapter \ref{ch:ch2} introduces canonical pulsars and millisecond pulsars (subcategories of radio pulsars) in detail with their observational properties and existing theoretical models. Chapter \ref{ch:ch3} provides detailed information on the datasets used and the formation of the profile parameter library of the dataset used in this work. Chapter \ref{ch:ch4} deals with the design, workflow, and application of the pulse signal simulator. Finally, chapter \ref{ch:ch5} presents relevant data analysis and the results, which suggests both the success and a few shortcomings of the thesis. 

%% file: Chapters/chap2.tex
Radio pulsars are rapidly rotating, highly magnetized neutron stars that emit beams of electromagnetic radiation from their magnetic poles. These beams sweep across the sky as the star rotates and, when aligned with the observer’s line of sight, are detected as periodic pulses. This phenomenon is commonly described by the \textit{lighthouse model} \citep{LorimerKramer2004}.

Pulsars typically have rotation periods ranging from milliseconds to a few seconds and possess strong magnetic fields of the order $10^{8} - 10^{14}$ G. Despite having masses comparable to the Sun, they are extremely compact objects with radii of only about 10–20 km. The observed pulsed emission is powered by the loss of rotational energy, as the neutron star gradually spins down due to electromagnetic torque. 

Canonical or normal pulsars are young radio pulsars with periods ranging from tens of milliseconds to seconds. They have very strong magnetic fields ($10^{11} - 10^{13}$ G) and thus have spin down rates ($\dot{P}=dP/dt$) in the range of $10^{-13}-10^{-16}$s/s. 

Millisecond pulsars (MSPs) form a distinct subclass of radio pulsars with much shorter periods ($P \sim 1$–$30$ ms) and weaker magnetic fields. They are believed to have been spun up through accretion in binary systems, resulting in highly stable rotation. This exceptional stability makes MSPs ideal tools for high-precision timing experiments.

\section{Observational Properties of Radio Pulsars}
The various properties of the observed emission from pulsars are explained such that the observed radiation is produced by the acceleration of charged particles along the field lines of highly magnetized rotating neutron stars (\cite{gold1968}; \cite{pacini1968}).

\subsection{Coordinates of observation}
Pulsars are generally found along the Galactic plane. Their observation coordinates are typically given in the equatorial coordinate system that is Right Ascension (RA) and Declination (Dec), because it remains nearly constant despite Earth's rotation. RA is analogous to longitude and is measured eastward along the celestial equator. It is traditionally measured in hours, minutes and seconds, where 24 hours is equivalent to 360 degrees. Dec is the celestial equivalent of latitude, measured in degrees North (+) and South (-) of the celestial equator. 

Some observations of pulsars are also made in the ecliptic coordinate system. This is based on the ecliptic plane, plane of Earth's orbit around the Sun. Ecliptic longitude (ELONG) is measured along the ecliptic eastward from the vernal equinox. The Sun crosses the celestial equator twice in a year, once moving northward along the ecliptic and later moving to the south. The point of intersection is called the vernal equinox and the autumnal equinox, respectively. Ecliptic latitude (ELAT) is measured perpendicular to the ecliptic plane. 

Because of a phenomenon called precession--a slow wobble of Earth's rotation axis caused by gravitational interactions with the Sun and Moon--the positions of the celestial equator and the vernal equinox shift slowly over time. To account for this shift, a \textbf{reference date} or an \textbf{epoch} must be specified when listing coordinates. \textbf{J2000} is the standard epoch used for astronomical catalogs, referring to the position of celestial objects at noon in Greenwich, England, on January 1, 2000. The `J' stands for the Julian epoch. In this work, J2000 coordinates are the standard input for defining the pulsar's position \citep{LorimerKramer2004}.

\subsection{Pulse period and period derivative}
Pulsars are characterized by their spin period. The rotation periods ranging from milliseconds to a few seconds suggest their origins. The short-period `millisecond pulsars' form a separate (older) population with different evolutionary histories from the long-period `normal pulsars'. 

But this period is not fixed. Astronomers observed that the pulsars are slowing down and usually at a very consistent rate. This `spin-down' is thought to be due to braking caused by rotating pulsar's magnetic field (\href{https://astronomy.swin.edu.au/cosmos/*/Pulsar+Characteristic+Age}{Swinburne University of Technology}). This information is also used to determine the characteristic age of the pulsar which is defined as :
\begin{equation}
    \tau = \frac{P}{2\dot{P}} = \frac{P}{2 dP/dt}
\end{equation}

where, P is the pulsar's period and $\dot{P}$ represents the period derivative (the rate at which the pulsar is slowing). 

\subsection{Integrated Pulse Profile} 
Because pulsars are intrinsically weak radio sources, only from the strongest sources individual pulses can be discernible above the noise. To improve the signal-to-noise ratio, individual pulses are time-averaged over a few thousand rotations through \textit{folding} and are coherently added, producing an \textbf{integrated pulse profile}. Although individual pulses may vary in shape, the integrated profile is remarkably stable at a given frequency, acting like a `\textit{fingerprint} of the neutron star's emission beam' as termed in \citep{LorimerKramer2004}.

\subsection{Interstellar Medium Propagation effects}
\begin{enumerate}
     \item \textbf{Dispersion} : Pulses observed at higher frequencies arrive earlier at the telescope than their lower frequency counterparts. \cite{hewish1968} - this effect is due to the frequency dependence of the group velocity of radio waves as they propagate through the ionized component of the ISM. Delay in pulse time arrival is inversely proportional to the observing frequency. The constant of proportionality, known as the dispersion measure (DM), is the integrated column density of free electrons along the line of sight. $$ DM = \int_{0}^{d} n_{e}dl $$
     where, d is the distance between the pulsar and Earth. 
     
     \item \textbf{Scintillation} : ISM is also highly turbulent and inhomogeneous, in addition to being magnetized and ionized. These irregularities produce phase modulations on the propagating pulsar signal that cause the observed intensity to fluctuate on a variety of bandwidths and timescales. This interstellar scintillation, first observed by \cite{lyne_rickett1968} is similar to optical 'twinkling' of stars caused by atmosphere of the Earth. \cite{scheuer1968} modeled the turbulent ISM as a thin screen of irregularities midway between the Earth and the pulsar. By considering the phase perturbations produced by such a screen, he demonstrated that the intensity fluctuations should be correlated over a characteristic scintillation bandwidth $\Delta f \propto f^{4}$, where f is the observing frequency. Scintillation bandwidth is the frequency scale over which pulsar signals remain correlated before scattering destroys coherence.

     \item \textbf{Pulse Scattering} : Due to the ionized interstellar medium (ISM), the scattered rays undergo a delay in their arrival, which combines to broaden an intrinsically sharp pulse profile. Scattering time decreases when the pulsar is observed at higher frequencies. The thin-screen model describes how the ISM affects the signal by treating the medium as an infinitely extended, very thin screen. The time delay follows inverse relation with the bandwidth as $\tau_{s} \propto 1/\Delta f \propto f^{-4}$. As explained in \cite{LorimerKramer2004}, this model can predict the broadening as the convolution of the true pulse shape with a one-sided exponential with a 1/e time constant known as the \textit{scattering time}, $\tau_{sc}$. Measurements of $\tau_{sc}$ for a large sample of pulsars show that it is correlated strongly with DM. More distant pulsars with larger DMs are more likely to be scattered. Undesirable effect when searching for pulsars since the scattering effectively stretches the true pulse shape which results in a reduction in the S/N ratio. Scattering time decreases when the pulsar is observed at higher frequencies. From the thin-screen model, $\tau_{s} \propto 1/\Delta f \propto f^{-4}$. Strong inverse frequency dependence of scattering favors searches carried out at high frequencies. 
     \\Scaling comes from the theory of Kolmogorov turbulence in the ISM.  
 \end{enumerate}

\subsection{Pulsar Magnetosphere}
In the beginning of pulsar magnetosphere research, \citep{gunn1970} considered magnetic dipole at the centre of the the star. The rotating dipole emits magnetic dipole radiation in the form of electromagnetic waves, which carries energy and momentum. Since neutron stars are perfect conductors, this induces an electric field (\textbf{$\Omega$} $\times \textbf{r}) \times \textbf{B}$, here \textbf{$\Omega$} is the rotation rate of the pulsar $\Omega=2\pi/P$ with P as its period. The electric field is balanced by the distribution of charge, which gives \textbf{E} a force-free condition inside the star. 
\begin{equation}
    \textbf{E} + \frac{1}{c}(\textbf{$\Omega$} \times \textbf{r}) \times \textbf{B})
\end{equation}

Due to the enormous electric force, charged particles on the surface will be stripped from it, move along the magnetic field line, and fill the magnetosphere. This procedure will create a plasma-filled magnetosphere and the plasma will co-rotate with the star. However, this co-rotation is maintained up to a certain distance which is known as the \textit{light cylinder}. This is because at this point the speed of plasma reaches the speed of light and it splits the magnetic field line into two regions: (a)\textbf{closed field lines}: that are within the light cylinder, enclosed by the light cylinder and (b)\textbf{open field lines}: that are outside the light cylinder. The last open field line is responsible for the emission cone and the region from where these lines emanate is the polar cap region of the pulsar.
\begin{figure}[h]
    \centering
    \includegraphics[width=0.5\linewidth]{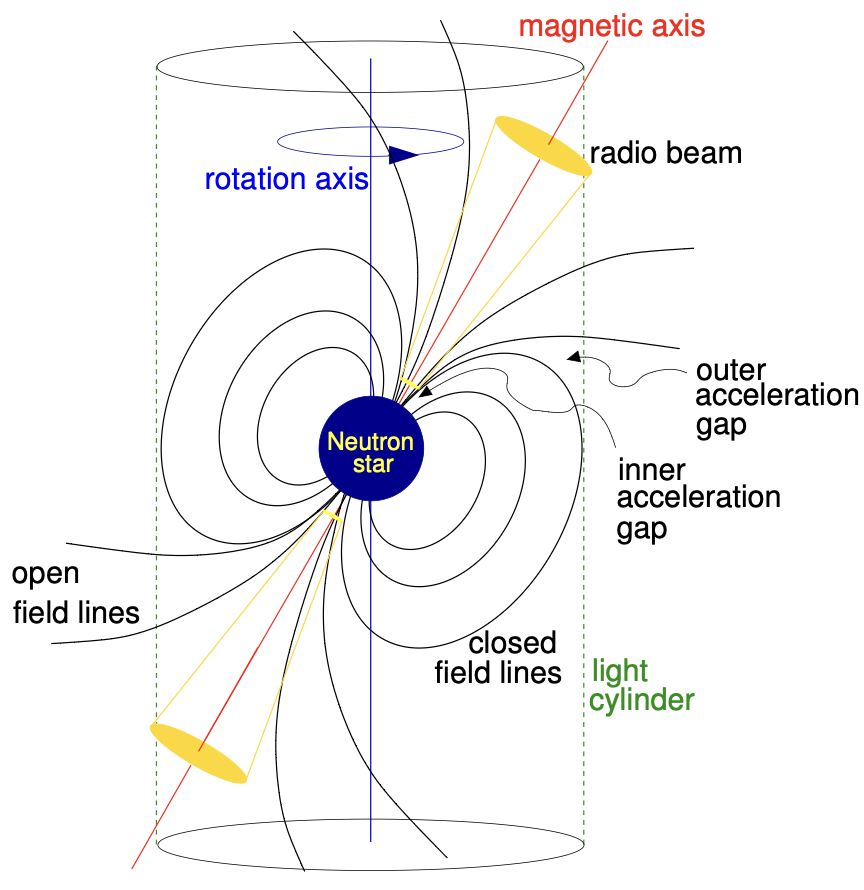}
    \caption{A toy model given by \cite{LorimerKramer2004} for the neutron star rotating and its magnetosphere}
    \label{fig:2.1}
\end{figure}

\section{Theoretical Models}
\subsection{Hollow-Cone emission model}
The hollow cone model proposed by \citep{RadhakrishnanCooke1969}, an empirical theory that explains the radiation beam of the pulsar that emanates from the magnetic poles. The radiation is not emitted uniformly from the poles, but forms a cone. Its structure can be understood to be `hollow', as there is no radiation at the magnetic pole. This is an important component of the rotating lighthouse model. As the pulsar rotates, this cone sweeps past the line of sight and the characteristic periodic, rapid pulses are observed.  

\begin{figure}[h]
    \centering
    \includegraphics[width=0.5\linewidth]{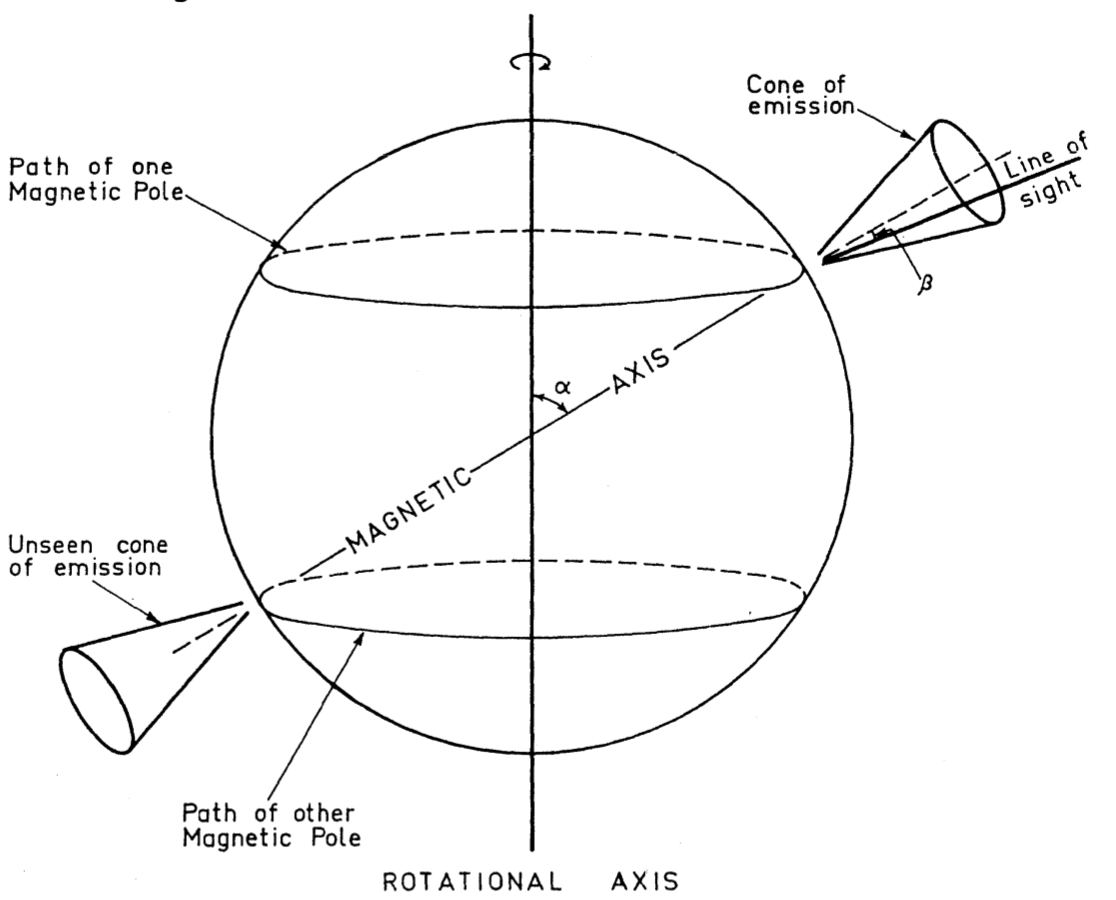}
    \caption{Geometry of pulsar emission beam \citep{RadhakrishnanCooke1969}}
    \label{fig:2.2}
\end{figure}

The observed pulse width is dependent on where the sight line of the observer cuts the emission beam cone. The geometrical structure of the pulsar's beam is defined by conical beam. The half-opening angle ($\rho$) of this cone is dependent on the extent of the open magnetic field region at the point of emission. This emission cone, which possesses an angular radius of $\rho$, is centered upon the magnetic axis. The magnetic axis is tilted relative to the rotation axis by an angle $\alpha$. Due to the rotation of the pulsar, an observer's view sweeps along a curved trajectory through the radiation beam. The segment of emission intercepted along this trajectory constitutes the observed pulse profile, and the length of that path as the pulse width. Furthermore, the minimum separation between observer's line of sight and the magnetic axis is quantified by the angle $\beta$.

\subsection{Core-Conal Emission Model}
The emission cone is theorized to be segregated into multiple, perhaps two, separate quasi-annular zones by \citep{Rankin1983}. Spectral differences exist between these zones: the conal emission has a comparatively flatter spectrum than the core radiation. Consequently, the core components primarily shape the pulse profile at lower observing frequencies, while the conal components become dominant at higher frequencies. The particular form of the profile is contingent upon how the observer's line of sight intersects the emission cone.  
\citep{Rankin1983}'s classification of the profiles: 

\subsubsection{Single Profile}
This is also divided into two categories based on the height of the emission : (1) conal single, and (2) core single profiles. \textbf{Conal single profiles ($S_{d}$)}, is attributed to a highly non-central trajectory. At lower frequencies, these profiles broaden and eventually bifurcate as the emission cone becomes larger. \citep{Rankin1983_II} depicts that at higher frequency, pulse widths follow a clean power law $f^{\sim -0.25}$. However, towards meter wavelengths, there is a broad narrowing, attributed to absorption feature. For frequencies lower than this, the profile width returns to its high frequency extrapolation.  
\\\textbf{Core single profiles ($S_{t}$)} exhibit much more central line of sight trajectory. \citep{Rankin1983_II} presents that the spectral evolution of pulse width for core single is virtually identical to that of the conal single including the absorption feature overlaid on the power-law dependence. As the observing frequency increases, pulse profiles evolve towards tripartite structure; at still higher frequencies, conal emission can become predominant, leading to a well-resolved double profile.  

\subsubsection{Tripartite Profile (T)}
The central component, resulting from nearly central cuts of the observer’s line of sight across the polar cap, displays characteristics typical of a core-type single profile. The outer components correspond to relatively inner traverses through the hollow conal emission region. As described in \citep{Rankin1983} T profiles can appear either as broad, well resolved or as narrower and partially blended features. These evolve to double structures at high frequencies, while at low frequencies they resemble core-single profiles. The separation between conal components follow a frequency dependence of approximately $f^{\sim -0.25}$. In certain cases the core component may appear slightly offset and can merge with, or even coincide with, one of the outer components.

\subsubsection{Double Profile (D)}
Double profiles are distinguished by strong conal components accompanied by broad, low-intensity, emission `bridge' in which weak core radiation may be present. When a clear signature of core emission is visible, they appear as `triple' profile \citep{Rankin1983}.
The separation between the unresolved conal components generally decreases with increasing observing frequency—showing a gradual decline above 1 GHz and a steeper trend at lower frequencies—although certain sources deviate from this overall pattern \citep{gunn1970} (\citep{Rankin1983_II}). They portray shallow absorption features. Their widths have more in common with the conal single whereas, well-resolved double profile represent highly central trajectories. With decrease in frequency, the widths broaden and it closely follows the variation seen in component spacing measurements.  

\subsubsection{Multiple Profile (M)}
Multiple profiles consists of five distinct components, incorporating both a central core component and contributions from inner and outer conal emission zones. The prominence of core feature varies from source to source, though it appears to be a common characteristic, likely resulting from the line of sight passing very near to the magnetic axis. With decreasing frequencies, M profiles generally evolve to broader triples (separating as $f^{\sim -0.25}$) and may further simplify into core-single profiles\citep{Rankin1983}.

%% file: Chapters/chap3.tex
Due to its extremely stable rotation over long periods, pulsars act as natural cosmic clocks. In particular, \textbf{millisecond pulsars (MSPs)} discovered by \cite{backer1982} are found in binary systems, attain their rapid rotation through the transfer of orbital angular momentum from a companion star \citep{RadhakrishnanSrinivasan1982}. Since MSPs are old pulsars that have been spun up by accretion, their earlier magnetic fields have been buried producing low magnetic field pulsars and spin down very slowly \citep{Alpar1982}. This stability allows them to serve as precise cosmic clocks through the technique of pulsar timing \citep{LorimerKramer2004}. 

\section{Pulsar timing}
Pulsar timing involves the precise measurement of the times of arrival (ToAs) of pulses emitted by a pulsar \citep{LorimerKramer2004}. These ToAs are compared against a predictive timing model to study variations in the pulsar's rotation and propagation effects. Due to their precise timing, MSPs are used to search for gravitational waves (GWs), which are light-speed ripples in the curvature of space-time that are caused by the extremely energetic processes of massive, accelerating objects in the universe  like the merging of compact objects like Neutron Stars (NSs) or Black Holes (BHs) (\citep{sathyaprakash2009}).

Laser Interferometer Gravitational-Wave Observatory (LIGO) observed GWs for the first time. While ground-based detectors like LIGO are sensitive to high-frequency GWs, an array of precisely timed MSPs can be used to detect low-frequency (nanohertz) gravitational waves. This technique treats individual MSPs as arms of the interferometer with respect to the Earth. This nanohertz detector, called the Pulsar Timing Array (PTA), relies on measuring minute changes in the times of arrival (ToA) of the radio pulses from the individual MSPs at the earth, as the passing GWs perturb the space-time between the pulsar and the earth \citep{rana2025inpta_dr2}. 

To further support the efforts of pulsar timing arrays, a realistic simulator for the pulse profiles of millisecond pulsars (MSPs) is essential. Such simulations require accurate input in the form of a well-defined library of pulse profile parameters, including component widths, amplitudes, and separations. However, currently there is no comprehensive, publicly available library of these parameters, which presents a significant gap for realistic simulation efforts. This work addresses this gap by constructing a dedicated library of profile parameters using the InPTA DR2 dataset, supplemented by profiles from the EPN database where necessary. This library is then used as input for generating realistic simulated pulse profiles.

\section{InPTA dataset}
The data used in this thesis are obtained from the Indian Pulsar Timing Array (InPTA: \citep{joshi2018inpta, tarafdar2022inpta}) Data Release 2 \citep{rana2025inpta_dr2}. InPTA observations are carried out using the upgraded Giant Metrewave Radio Telescope (uGMRT), an interferometer with 30 antennas, each with a diameter of 45 m spread out over a 25 km area, the largest telescope at this lower frequency range. Fourteen of these antennas are located in a central square, while the remaining twelve antennas are distributed along three arms in a `Y' shape. The uGMRT provides four observing bands: band 2 (120-250 MHz), band 3 (250-500 MHz), band 4 (550-850 MHz), and band 5 (1050-1450 MHz). The InPTA observations are carried out by splitting the uGMRT antennas into multiple subarrays, observing the same source in different frequency bands simultaneously, with 100 MHz or 200 MHz bandwidth in each band depending on the observing epoch \citep{rana2025inpta_dr2}. 

\subsection{Data Release 2 Observation}
The InPTA DR2 includes 28 MSPs with observing time spanning 7.5 years covering the period between 2016-2024. In observation cycle 45, used in this work, InPTA has followed a hybrid observing strategy where alternate observing sessions are held in band 3-only (only one subarray) or band 3 + band 5 (two subarrays) configurations with 200 MHz bandwidth in each band. This adjustment allowed to include more pulsars in the band 3-only observation mode in this period (2022-2024) by utilizing more uGMRT antennas in a single subarray, and hence significantly increasing sensitivity and enabling pulsar detection within shorter observation periods \citep{rana2025inpta_dr2}. 

The data includes the following pulsars: J0030+0451, J0034-0534, J0437-4715, J0610-2100, J0613-0200, J0614-3329, J0740+6620, J0751+1807, J0900-3144, J1012+5307, J1022+1001, J1125+7819, J1545-4550, J1600-3053, J1643-1224, J1705-1903, J1713+0740, J1730-2304, J1744+0740, J1745+1017, J1857+0943, J1909-3744, J1910+1256, J1939+2134, J1944+0907, J2124-3358, J2145-0750, J2302+4442. 

\begin{table}[h!]
\centering
\footnotesize
\caption{Summary of observing parameters for InPTA data.}

\begin{tabularx}{\textwidth}{|c|c|c|c|c|X|c|c|X|}
\hline
\textbf{Obs.} & \textbf{No.} & \textbf{MJD} & \textbf{MJD} & \textbf{Band} & \textbf{Frequency Band (MHz)} & \textbf{Channels} & \textbf{Sampling ($\mu$s)} \\
\textbf{Cycle} & \textbf{PSRs} & \textbf{Start} & \textbf{End} &  &  &  &  \\
\hline

45 & 28 & 60245 & 60399 & 3 & 300--500 & 128 & 5.12 \\
   &    &       &       & 5 & 1260--1460 & 1024 & 40.96 \\

\hline
\end{tabularx}
\end{table}

\subsection{Data Processing Overview}
The uGMRT data are converted into a standard pulsar archive format within the InPTA processing pipeline. The uGMRT beamformed data are initially recorded in filterbank format and processed within the InPTA pulsar data analysis pipeline. The processing includes radio frequency interference (RFI) mitigation, de-dispersion to correct for interstellar dispersion, and folding using known pulsar ephemerides to generate integrated pulse profiles.

RFI mitigation and data format conversion are performed using \texttt{RFIClean}, preserving intrinsic signal characteristics \citep{Maan2021RFIClean}. The processed data are stored in the standard FITS (Flexible Image Transport System)-based format for pulsar data files, \texttt{PSRFITS} format \citep{Hotan2004PSRFITS}. De-dispersion and folding are carried out using tools such as \texttt{dspsr} \citep{vanStraten2011}. The final outputs are integrated pulse profiles (e.g., \texttt{rficlean.fits}), which are used for further analysis.

Since this study focuses on pulse profile analysis and parameter extraction, all subsequent analysis is performed directly on these processed \texttt{PSRFITS} files provided by the InPTA collaboration.

\begin{figure}[H]
    \centering
    \includegraphics[width=0.8\linewidth]{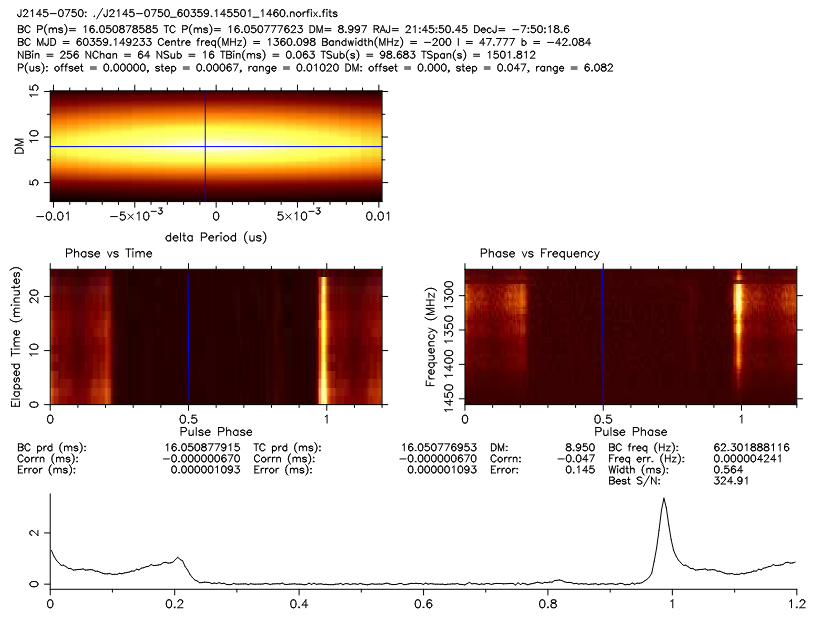}
    \caption{Data obtained from InPTA dataset}
    \label{fig:3.1}
\end{figure}

\section{EPN Database}
The European Pulsar Network (\href{https://psrweb.jb.man.ac.uk/epndb/about.html}{EPN}) Database of Pulsar Profiles, provided by the University of Manchester, is a comprehensive collection of pulsar profiles presented in a standardized format. It includes profiles for over 1000 pulsars obtained from a variety of observational campaigns, along with the corresponding raw data in both ASCII and EPN formats (\href{https://www.jb.man.ac.uk/pulsar/Resources/epn/}{EPN resources}). 

These observations were conducted using the Parkes Observatory (Murriyang), a 64-m radio telescope operated as part of CSIRO's Australian Telescope National Facility (ATNF). Its large collecting area allows for high-sensitivity observations, making it well-suited for detailed pulse profile studies.

For a subset of MSPs in this work, where the InPTA data were not suitable for reliable parameter extraction, pulse profile data in \texttt{PSRFITS} format were obtained from the EPN database. In particular, profiles from the multifrequency polarization study of MSPs by \cite{Dai2015} were used. The EPN data utilized in this work were that have been both time- and frequency-scrunched, resulting in integrated pulse profiles. These were directly used for parameter extraction, ensuring consistency in analysis methodology with the InPTA data. 

\begin{figure}[H]
    \centering
    \includegraphics[width=0.9\linewidth]{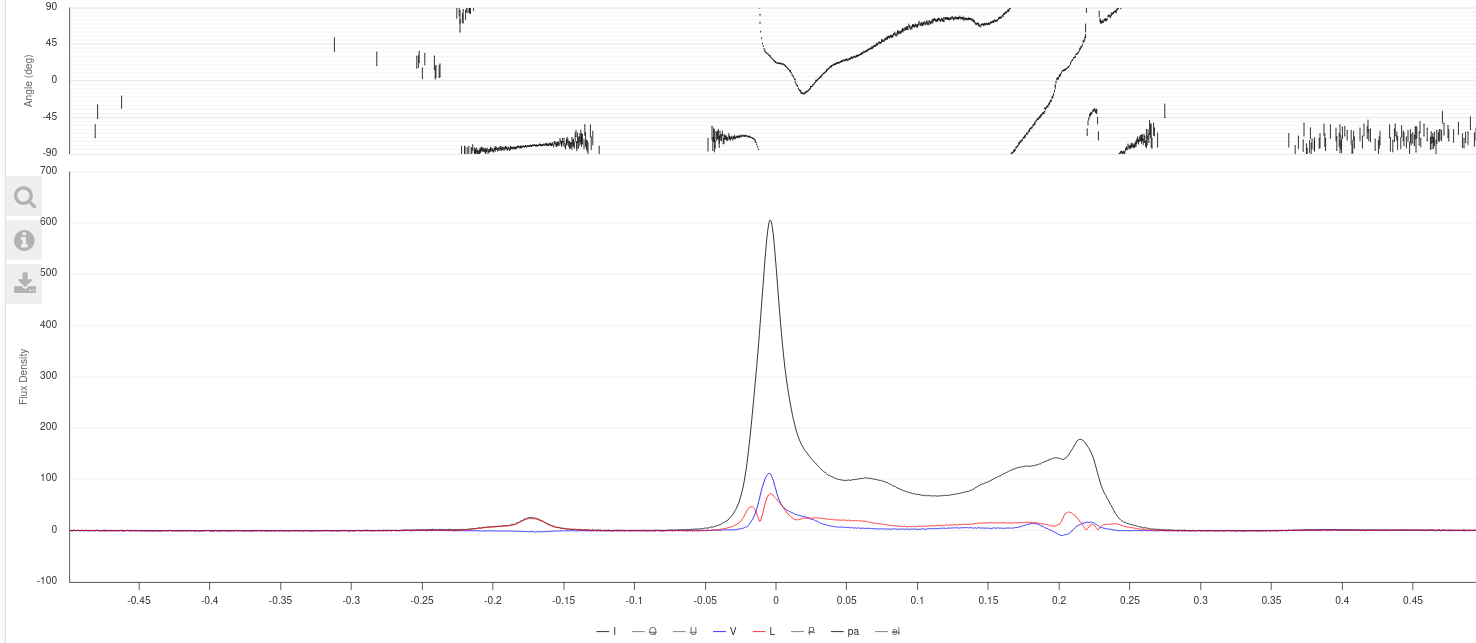}
    \caption{EPN database example data}
    \label{fig:3.2}
\end{figure}

\section{Profile Parameter Extraction}
To generate a library of the profile parameters of the InPTA's 28 millisecond pulsars (MSPs), an original python-based analytical pipeline designed to automate the extraction of the gaussian parameters was developed. J1705-1903 had a very low S/N and thus is not utilized for the generation of this parameter library. 
For the rest 27 MSPs, iterative least-square minimization technique and statistical model selection is carried out to provide accurate inputs for a realistic pulse profile simulator. 

The analysis involves a structured pipeline developed to initially process the data, estimate noise, and optimize the multi-component gaussian fitting. 

Gaussian components can be considered a flexible model to interpret the physical properties of the pulsars. They have been commonly applied to model the different emission components of pulsars due to the distinct regions of the magnetosphere. This model can be effectively applied to compare the complex profiles with a few parameters.

\subsection{Data Preparation}
The initial stage of the pipeline involves preparing the pulsar integrated profile for fitting. Pulse profiles are loaded in \texttt{PSRFITS} format. A preprocessing step is first performed to ensure compatibility with the downstream analysis. The \texttt{TELESCOP} header name is modified to `GM' (GMRT convention) while preserving the original data. 

For InPTA, Band 5 data has been utilized, as at higher frequencies there are minimum interstellar medium effects to alter the intrinsic profile properties. Hence, it is ideal as an input. This is complemented with the profiles from EPN database at 1369 MHz. Where InPTA profile data have low signal-to-noise ratio and produce inadequate fitting, the EPN database has been used to fulfill the purpose, as we require \textit{realistic} input for our simulator. 

The preliminary data processing and reduction for the profiles are conducted using \textbf{\texttt{psrchive}}, an open source C++ development library for the analysis of pulsar astronomical data. Described by \cite{Hotan2004PSRFITS}, the library provides a robust suite of algorithms for tasks such as pulsar timing, scintillation studies, and RFI mitigation.  In this analysis, \texttt{psrchive} is used to interface with pulsar data stored in the \textbf{PSRFITS} format. 

The analysis pipeline utilizes specific \texttt{PSRCHIVE} methods to integrate the raw observation data into a high signal-to-noise ratio profile. This includes `scrunching' multi-dimensional data by :
\begin{enumerate}
    \item \textbf{tscrunch()} : integrating all time sub-integrations into a single profile. 
    \item \textbf{fscrunch()} : summing all frequency channels to provide a frequency-integrated profile. 
    \item \textbf{pscrunch()} : summing the polarization information to produce a total intensity profile.
\end{enumerate}
This produces a single one-dimensional pulse profile. 
\texttt{ar.get$\_$data}, a \texttt{psrchive } command is then used to generate a 4D array [subint, pol, chan, bin] which selects the actual pulse profile array over the bins. 

\subsection{Baseline and noise estimation}
To ensure that the fitting algorithm is numerically stable, a rough centering of the profile is performed by identifying the peak bin and rolling the profile so that the primary peak is centered at phase 0 (with the phase going from -0.5 to +0.5). 

A robust definition of the baseline and noise level ($\sigma_{noise}$) is essential for accurate parameter estimation. The \textit{off-pulse region} is defined from the outer 25$\%$ of the profile (which may be adjusted depending on the profile morphology, particularly at the edges). The baseline is calculated as the median of this region as it picks the middle value which is robust against the outliers. Baseline estimates the system level when the pulsar is not emitting and it is typically understood as the background intensity of the profile. $\sigma_{noise}$ is estimated using a robust statistical approach, \textbf{Median Absolute Deviation (MAD)}: 
$$ \text{MAD} = \text{median}(|x_{i}-\text{median}(x)|) $$
$$ \sigma_{noise} = 1.4826 \times \text{MAD} $$
$$ \implies \sigma_{noise} = 1.4286 \times \text{median}(|prof_{original}[off]-\text{baseline}|) $$

where, $prof_{original}[off]$ is the off-pulse region of the profile defined and the baseline is the median of this off-pulse region as defined above. This scaling ascertains that the normally distributed noise, the estimate remains consistent with the standard deviation \cite{berendsen2011data}. 

The profile is then baseline-subtracted to obtain a zero-level corrected signal :
$$ profile = prof_{original} - baseline $$

The pulse window is then defined to restrict the fitting region. A threshold of 1-8$\%$ of the peak amplitude is used depending on the amplitude of the smallest component of the profile. The corresponding phase range defines the region containing significant emission. This avoids fitting noise-dominated regions. 

\subsection{Multi-component Gaussian model}
The integrated profile is modeled as a summation of N circular Gaussian components to account for the periodic nature of the pulsar phase. Each component is defined by its amplitude ($A$), mean phase ($\mu$), and width ($\sigma$) (not to confuse with noise $\sigma_{noise}$) : 

$$ g_i(x) = A_i \exp\left(-\frac{\Delta x^2}{2\sigma_i^2}\right) $$
$$ I(x) = \sum_{i=1}^{N} g_i(x) $$

where $\Delta x$ is the minimum distance between the phase x and the mean $\mu$, accounting for wrap-around at the phase boundaries. 

\subsection{Iterative Fitting and Model Selection}
A critical challenge in profile modeling is determining the optimal number of Gaussian components (N). The pipeline performs a scan across N (typically from 1 to 10). For each N, a nonlinear least-squares minimization is performed using \texttt{lmfit} package to minimize the sum of squared differences (residuals) between the data ($y_{i}$) and the multi-component model $F(x_{i})$: 
$$ Residuals \text{ = } \frac{y_{i}-F(x_{i})}{\sigma_{noise}} $$
$$ \chi_{0}^{2} = min \left( \sum_{i=1}^{n} \frac{|| y_{i} - F(x_{i}) ||^{2}}{\sigma_{noise}^2} \right) $$ 
where weighted least square minimization of the residuals has been carried out. 

To determine the `best' N (the number of Gaussian components that accurately describe the physical emission without capturing stochastic noise), the pipeline evaluates models ranging from N=1 to N=10 using two primary metrics : 
\begin{enumerate}
    \item \textbf{RMS Residuals :} A measure of the average deviation of the fit from the observed data \citep{berendsen2011data}. In the analytical pipeline, it is calculated as : 
    $$ RMS = \sqrt{\frac{1}{n} \sum_{i=1}^{n} (y_i - F(x_i))^2} $$
    While a decreasing RMS typically indicates a better fit, it does not inherently account for the `cost' of adding parameters, which can lead to overfitting. 
    
    \item \textbf{Reduced $\chi^{2}$ ($\chi_{red}^{2}$) :} This provides a more statistically robust criterion by weighting the squared deviations by the estimated noise levels ($\sigma_{noise}$) and normalizing by the \textbf{degrees of freedom ($\nu$)} \citep{berendsen2011data}. 
    $$ \chi_{red}^{2} = \frac{\chi_{0}^{2}}{\nu} $$
    \begin{itemize}
        \item \textbf{Degrees of freedom ($\nu$) :} This is defined as $\nu=n-m$, where n is the number of data points and m is the number of adjustable parameters. For a fit with N Gaussian parameters, each component contributes three parameters (amplitude, mean, and width), making m=3N. 

        \item \textbf{Selection Logic :} The pipeline identifies the best `N' by minimizing the absolute difference between $\chi_{red}^2$ and 1. 
        \begin{itemize}
            \item $\chi_{red}^2 \sim 1$ : Indicates that the residuals are consistent with the random noise level of the data, signifying a statistically justified fit.
            \item $\chi_{red}^2 \gg 1$ : Suggests that the model is under-fitted; the Gaussian components are insufficient to capture all significant emission features. 
            \item $\chi_{red}^2 \ll 1$ :  Suggests that the model is over-fitted or that the noise level ($\sigma$) was overestimated. In this case, the model may be treating random fluctuations as physical components.
        \end{itemize}
    \end{itemize}
\end{enumerate}

\begin{figure}[H]
    \centering

    \begin{subfigure}{0.48\textwidth}
        \centering
        \includegraphics[width=\linewidth]{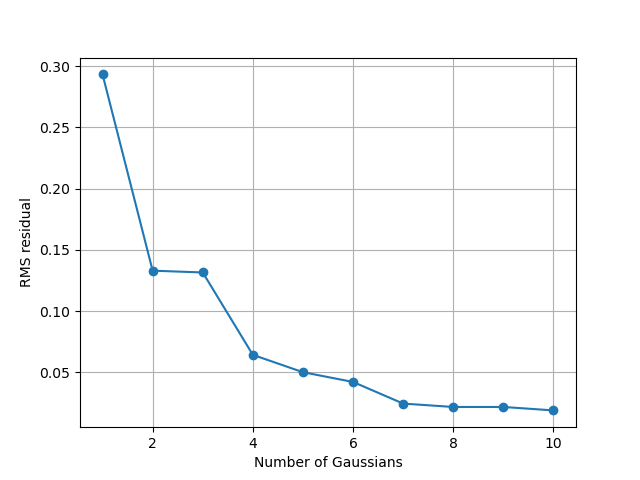}
    \end{subfigure}
\hfill
    \begin{subfigure}{0.48\textwidth}
        \centering
        \includegraphics[width=\linewidth]{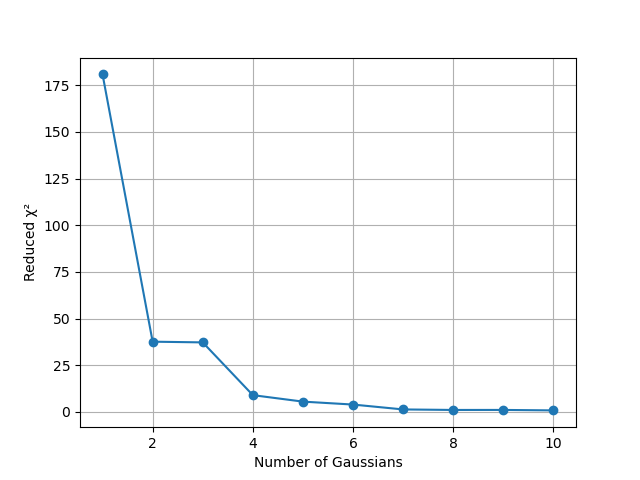}
    \end{subfigure}
    
\caption{RMS and Reduced chi square for number of components selection for PSR J2145+0750} 
\label{fig:3.3}
\end{figure}

\subsection{Parameter Extraction for Simulation}
Once the optimal $n_{best}$ is determined, a final least-square minimization is performed to extract the analytical parameters required for the pulsar simulator. 

\begin{enumerate}
    \item \textbf{Amplitude (A) } 
    
    The amplitude represents the peak flux intensity of an individual Gaussian component. In the multi-component model, the total flux at any phase is the summation of these individual peaks weighted by their respective Gaussian distributions. 

    \item \textbf{Phase ($\mu$): Centering and Alignment} 

    The mean phase position ($\mu$) of each component is extracted and re-aligned to the original data's phase space. This involves a two-step process:
    \begin{itemize}
        \item Initial Centering : To improve fitting stability, the raw profile is rolled so that the primary peak is at phase 0, creating a \texttt{phase$\_$shift}. 
        \item After fitting, the simulated phase is corrected by subtracting the initial shift and applying a modulo operation: $phase\_sim = (mu - phase\_shift) \% 1.0$. This ensures the simulation accurately reflects the observed rotational phase of the pulsar. The phase is defined from 0 to 1 in the simulator, so this correction is required for that as well. 
    \end{itemize}

    \item \textbf{Full Width at Half Maximum ($w_{50}$) :} 

    While the fitting algorithm optimizes the Gaussian standard deviation ($\sigma$), the simulator requires the FWHM, denoted as $w_{50}$. This is the width of the pulse at half of its peak amplitude. For a Gaussian distribution, $w_{50}$ is derived using the constant: $ w_{50} = 2.355 \times \sigma $
 
\end{enumerate}

\begin{figure}[H]
    \centering
    \begin{subfigure}{0.5\textwidth}
        \centering
        \includegraphics[width=0.9\linewidth]{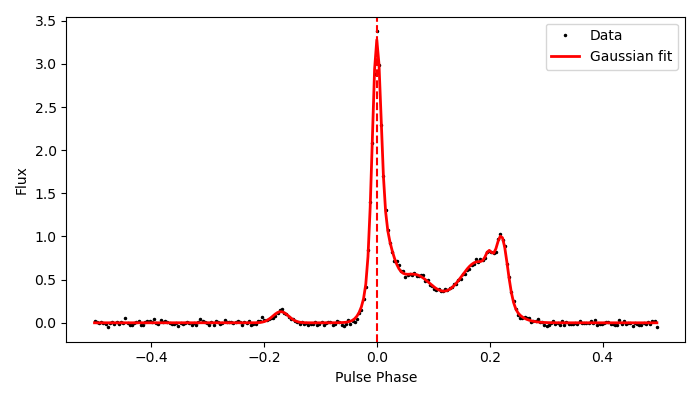}
        \caption{Final Fitting (centered)}
        \label{fig:3.4(a)}
    \end{subfigure}%
    \begin{subfigure}{0.5\textwidth}
        \centering
        \includegraphics[width=0.9\linewidth]{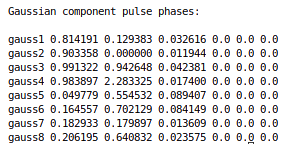}
        \caption{Profile Parameters (center shifted)}
        \label{fig:3.4(b)}
    \end{subfigure}
    \caption{Fitting for PSR J2145-0750}
    \label{fig:3.4}
\end{figure}

Though the application of the Gaussian model is quite effective to interpret the profiles of the pulsars, the application of the model is not unique. Moreover, the identification of the components of the millisecond pulsars with complex and blended emission profiles might be ambiguous. However, the application of the statistical model selection criterion is quite effective to overcome the limitations of the model.

\begin{figure}[H]
    \centering
    \includegraphics[width=0.5\linewidth]{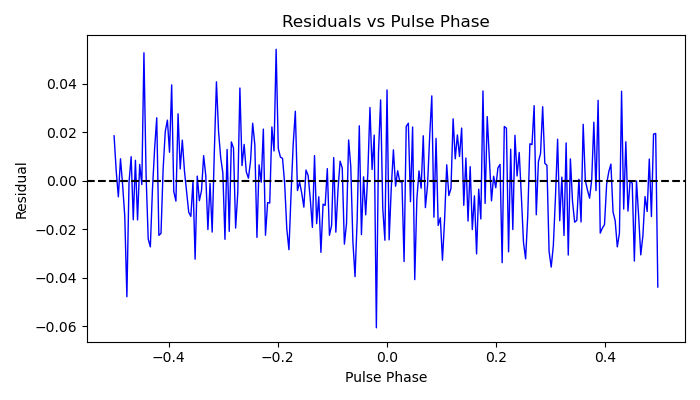}
    \caption{Residual (Data-Model) : Random behavior around 1 indicates that the model is statistically correct}
    \label{fig:3.6}
\end{figure}

%% file: Chapters/chap4.tex
This chapter presents the primary contribution of this work that was to develop a simulation framework for generating realistic integrated profiles of pulsars. This software has been developed on a monolithic python code written by \textbf{MA Krishnakumar}, Reader-F at Radio Astronomy Centre (RAC-TIFR), Ooty. This is carried out by designing a modular python package \texttt{\textbf{pulsar$\_$sim}} to simulate pulsar observations with high physical fidelity. The simulator serves as a bridge between analytical models--such as the Gaussian parameters extracted and synthetic data archives that can be processed using standard pulsar analysis software like \textbf{\texttt{psrchive}}. 

The motivation for constructing a pipeline like this arises from the need to comprehensively study pulse profile evolution under controllable conditions, mainly in the presence of interstellar medium (ISM) effects, such as scattering and dispersion. 

Although the observational data (detailed in \ref{ch:ch3}) provide real measurements, they do not provide independent control over physical parameters such as dispersion measure (DM) fluctuations and scattering strength in time and do not allow us to understand the properties of these profiles in the frequency range that is not yet currently observable accurately. By leveraging the \textbf{\texttt{psrchive}} library, the simulator produces standardized \textbf{PSRFITS} outputs, ensuring full compatibility with existing astronomical analysis pipelines. 

\section{Simulation Design and Modular Architecture}
The simulation was designed with the following principles: 
\begin{itemize}
    \item \textbf{Modularity :} Each physical process (DM variation, scattering, intrinsic profile generation) is implemented independently. 
    \item \textbf{Configurability :} All parameters are externally controlled by command-line arguments and input configuration files. 
    \item \textbf{Physical realism :} Frequency scaling laws and stochastic processes are incorporated.
    \item \textbf{Compatibility :} Output conforms to \textbf{PSRFITS} standards using the \texttt{psrchive} library. 
\end{itemize}

Initially implemented as a monolithic script, the pipeline was later refactored into separate modules forming a structured software package. The specialized modules handle distinct physical or data-handling tasks : 

\begin{enumerate}
    \item \textbf{\texttt{cli}:} Handles the command-line interface and configuration parsing
    \item \textbf{\texttt{io}:} Interfaces with pulsar ephemeris (.par) files to retrieve positional and timing metadata. 
    \item \textbf{\texttt{dm$\_$series} and \texttt{scat$\_$series} :} Generate temporal stochastic realization of ISM-induced variations. 
    \item \textbf{\texttt{profiles}:} The core mathematical engine for profile synthesis and signal degradation
    \item \textbf{\texttt{profiles}:} An interface with the \texttt{psrchive} library to populate FITS data structure.
    \item \textbf{\texttt{utils}:} High-performance utilities applying Fast Fourier Transforms (FFTs) for circular convolution and power-law time series generation. 
\end{enumerate}

\section{Simulation Workflow and Initialization}
The simulation flow begins with the \texttt{main()} function, which controls the data pipeline. 

\subsection{Configuration and Ephemeris Parsing}
The simulator initializes by parsing terminal arguments through the \texttt{parse$\_$arguments()} module. Key parameters include the MJD range, that is the observation time span (\texttt{mjd$\_$start} to \texttt{mjd$\_$finish}), frequency range, number of subintegrations, number of channels (\texttt{nchan}), phase bins (\texttt{nbin}), and number of simulated epochs. The other parameters passed are the input file consisting of the fitting parameters, spectral index of the pulsar, flux density, peak signal-to-noise ratio (S/N) of each channel, and switches to control ISM effects. 

The simulator then calls \texttt{readparfile()} to load a pulsar ephemeris file containing the information about the observation of that pulsar. This module extracts : 
\begin{itemize}
    \item \textbf{PSRJ :} Pulsar name 
    \item \textbf{RAJ/DECJ :} Celestial coordinates of the pulsar. If only ecliptic coordinates (ELONG, ELAT) are present, the simulator utilizes the astropy library to transform them into the standard J2000 frame.
    \item \textbf{DM :} The reference Dispersion Measure.  
\end{itemize}

\subsection{Input File \texttt{simpulse$\_$input.txt}}
The \texttt{simpulse$\_$input.txt} file is the key input of this simulator, as it provides the number of components and their parameters for each specific pulsar. It defines the `intrinsic' pulse by listing the Gaussian and Lorentzian components, scattering characteristics, and dispersion variations, referenced to 1 GHz observing frequency. 

The previously extracted \textbf{Gaussian parameters}, mean phases, amplitudes, and widths are utilized here. Along with this, this file also provides the means to provide the frequency evolution parameters. Since it has been observed throughout these years by astronomers that MSPs show very little intrinsic frequency evolution in their profiles and do not follow a specific trend, these parameters have been set to 0.0. For normal pulsars, these parameters can be provided to simulate the evolution. 

The next function is the \textbf{scattering time reference and variability} \texttt{scat}. It defines the multi-path scattering effects of the ISM as the pulsar signal passes through it. The scattering time ($\tau_{sc}$) of the pulsars, also referred to as scattering timescale or pulse broadening time, represents the exponential tail added to a pulse profile by interstellar radio wave scattering. This value is highly frequency-dependent, generally following a power-law relation $\tau_{sc} \propto \nu^{\alpha} DM^{\beta}$, with the scattering index $\alpha$ usually around -4.4 and the default value for $\beta$ is -2.0, consistent with the Kolmogorov spectrum of the ISM. $\tau_{sc}$ (tscat) varies for each pulsar as it highly depends on its DM and position in the sky. Therefore, YMW16\footnote{https://www.atnf.csiro.au/research/pulsar/ymw16/} energy-density model \cite{Yao2017YMW16} is used to estimate $\tau_{sc}$ for each object. PyGEDM\footnote{https://github.com/FRBs/pygedm}, a python interface for the various electron-density model provides a web application, which was used for this purpose. The RA/DEC and DM for each pulsar were obtained from \textit{ATNF Pulsar Catalogue}\footnote{https://www.atnf.csiro.au/research/pulsar/psrcat/\label{atnf}} to find the value of $\tau_{sc}$ (usually in nanoseconds for MSPs). 

The default value of $\alpha$ is -4.4 but it may vary with each pulsar and this code also gives the option to pass the value if known as an argument. 

Options for different modes are available that govern how scattering evolves with different observing epochs. 

To simulate realistic temporal variations in the pulsar's dispersion measure (DM) across different MJDs, the \textbf{DM spectral model} \texttt{dmsp} defines a power-law spectral model defined by amplitude and spectral index (\texttt{spidx}). The simulator utlilizes the value: amplitude set to 1$\times 10^{6}$ as default and \texttt{spidx} of each pulsar obtained from the \textit{ATNF Pulsar Catalogue}\footref{atnf}.

The simulator also supports the injection of discrete, non-stochastic ISM events, such as an \textbf{exponential dip and recovery}. It is dependent on the epoch (MJD at which the event occurs), amplitude (magnitude of variation), recovery time (the time constant for the signal to return to its baseline), and the gaussian parameter of the specific component being affected. 

%in simpulse_input : scattering timescale (YMW16 model), dm of each pulsar, par files of each pulsar - ATNF pulsar catalogue, params
%residues of simulation and the data - compare 

\section{Temporal Modelling}
Before synthesizing individual profiles, the simulator generates a time series for parameter evolution across multiple epochs. 

\subsection*{DM time series}
DM is modeled as a speculative process with a power-law spectrum:
$$ P(f) \propto f^{\beta} $$ where beta is the spectral index of the pulsar. This is implemented by generating a Fourier-domain spectrum and transforming into a time series. DM variations are produced by \texttt{make$\_$dmseries} module, which then calls the \texttt{make$\_$timeseries} function to perform inverse FFT to produce a continuous, correlated series in the time domain. 
The amplitude and the spectral index are configurable in the input file for a realistic simulation of the pulsar. 

\subsection*{Discrete ISM events}
The simulator also allows for discrete ISM events or `dips' due to localized ISM perturbations through the function \texttt{make$\_$dips}. They are modeled using exponential or Gaussian recoveries, characterized by specific epoch, amplitude, and recovery time constant. 

\subsection*{Scattering series}
The scattering timescale, modeled as: $\tau_{sc} \propto \nu^{\alpha} DM^{\beta}$, defines how the scattering time evolves. There are four distinct modes that are supported in this module \texttt{make$\_$scatseries}:

\begin{enumerate}
    \item \textbf{stable:} Constant scattering for all epochs.
    \item \textbf{dm:} Scattering variations that scale identically to the Dispersion Measure (DM) time series.
    \item \textbf{random:} Scattering parameters vary stochastically based on the provided rms (root mean square) value.
    \item \textbf{powerlaw:} Uses the given spectral index to generate a scattering time series from a power spectrum.
\end{enumerate}
\texttt{params} can be used to specify which variable tscat/alpha is targeted by the selected variability mode. 

\section{Synthesis of Integrated Profile}
The core of the simulation occurs in the \texttt{make$\_$fits}, which calls \texttt{make$\_$profile} for every frequency channel and sub-integration. While \texttt{make$\_$profile} is responsible for mathematically constructing the pulse signal for a single frequency channel at a specific epoch, \texttt{make$\_$fits} orchestrates the entire observing system, integrating these profiles into a standardized astronomical data format.

\subsection{Intrinsic Pulse Construction}
The noise-free profile is built by summing individual mathematical components. 
\begin{itemize}
    \item \textbf{Gaussian Components:} Synthesized using: $$ g(x; A, \mu, \sigma) = A \times exp \left( -\frac{(x-\mu)^{2}}{2\sigma^{2}}  \right) $$ where $\sigma$ is calculated from the input $w_{50}$ (Full Width at Half Maximum) as $\sigma=w_{50}/2.35482$. 

    \item \textbf{Lorentzian Components:} Defined for broader emission wings: 
    $$ L(x; A, \mu, \sigma)=A \times \frac{\gamma^{2}}{\gamma^{2}+(x-x_{0}^{2})} $$
\end{itemize}

In this work, only Gaussian parameters have been extracted and are being used in the simulation but can be extended to Lorentzian and exponential dips in the profile. 

\subsection{Frequency Scaling and Flux Calibration}
The simulator applies physical scaling laws to ensure that the profile evolves realistically with frequency:
\begin{itemize}
    \item \textbf{Spectral index:} Flux density is scaled relative to a 1400 MHz reference using the spectral index $\beta$:
    $$ S_{freq} = S_{1400}(freq/1400)^{\beta} $$
    \item \textbf{Width Scaling:} Component widths are adjusted according to the power law frequency dependence defined in the input file. 
\end{itemize}

\section{Modeling ISM effects}
As the signal passes from the pulsar to Earth, it is affected in a number of ways (explained in \ref{ch:ch3}) through the interstellar medium. This is taken care of in this modeling for dispersion, DM delay, scattering, and DM smearing. 

\subsection{Dispersion delay}
Electromagnetic radiation from pulsar experiences a frequency-dependent index of refraction as they propagate through the ISM, a cold, ionised plasma. The refractive index is approximately given by \citep{LorimerKramer2004}:
\begin{equation} \label{eq1}
    \mu=\sqrt{ 1-\left( \frac{f_{p}}{f}\right)^2 }
\end{equation}

where f is the observing (wave) frequency and $f_{p}$ is the plasma frequency
\begin{equation}
    f_{p}=\sqrt \frac{e^{2} n_{e}}{\pi m_{e}}
\end{equation}
where, $n_{e}$ is the electron number density, e and $m_{e}$ are the charge and mass of an electron respectively. A wave will not propagate if $f<f_{p}$. 

From \ref{eq1}, $\mu<1$ which suggests that the group velocity of a propagating wave $v_{g}=c\mu$ is less than the speed of light. As a result, the propagation of a radio signal along a path of length \textbf{d} from pulsar to Earth will be delayed in time with respect to a signal of infinite frequency with an amount dependent of $f^{-2}$. The magnitude of this effect is quantified by Dispersion Measure (DM), representing the integrated column density of free electrons along the line of sight:
\begin{equation}
    DM = \int^{d}_{0} n_{e}dl
\end{equation}

Lower frequency radio waves travel slower than the higher frequency waves through the ionised medium. The relative time delay $(\Delta t)$ between two frequencies $f_{1}$ and $f_{2}$ (in GHz) is, 
\begin{equation}
    \Delta t = 4.15 \times 10^{6}\text{ ms } \times (f_{1}^{-2}-f_{2}^{-2}) \times \text{DM}
\end{equation}
which has also been incorporated in the simulator. 

\subsection{Scatter Broadening} 
The multi-path propagation of the radio signal due to the inhomogeneities and turbulence in the ISM plasma, scattering, is responsible for the same pulse taking slightly different paths to the observer. These paths have different lengths, leading to a geometric delay in the arrival of parts of the signal. 

This effect of scattering is modeled as the convolution of the intrinsic pulse profile with a \textbf{Pulse Broadening Function (PBF)}. A scattering exponential decay tail on the pulse at lower frequencies is apparent as a result of the FFT-based cirular convolution of profile with a PBF. Depending on the physical environment of ISM, four distinct \cite{Williamson1972} and \cite{Geyer2017} approximations are implemented in the simulator:

\begin{enumerate}
    \item \textbf{PBF1 (Thin screen):} A standard exponential decay, $$PBF(x)=exp(-x/\tau_{sc})$$
    \item \textbf{PBF2 (Thick screen):} Models scattering in an extended medium using: $$ PBF_{2}(x)=\sqrt \frac{\pi \tau_{sc}}{4x^{3}} \text{ exp} \left(-\frac{\pi^{2}\tau_{sc}}{16x} \right) $$
    \item \textbf{PBF3 (Truncated screen):} A specialized approximation that averages early bins to avoid numerical `zero bin' artifacts common in scattering simulations. 
    \item \textbf{PBF4 (Continuous media):} Models scattering in a uniform turbulent medium: $$ PBF_{4}(x)=\sqrt{\frac{\pi^{5}\tau_{sc}^{3}}{8x^{5}}} \text{ exp} \left( -\frac{\pi^{2}\tau_{sc}}{4x} \right) $$ 
\end{enumerate}

\subsection{DM smearing}
The amount of delay in the pulse depends on the frequency and DM. The amount of dispersion measure smearing $\tau_{DM}$ across a finite bandwidth ($\Delta f$) MHz at a central frequency $\nu$ GHz is proportional to DM is given by:
\begin{equation}
    \tau_{DM}=8.3 \times \frac{\Delta f}{\nu^{3}} \times \text{DM} \text{ } \mu s
\end{equation}
Within B, the arrival times vary across the channel, causing the pulse to appear broadened or `smeared'. The simulator accounts for this by applying \textbf{symmetric boxcar convolution}. This ensures that the pulse width of the component increases, preserving the position of the peak. The implementation follows the following steps:
\begin{enumerate}
    \item \textbf{Bin Conversion:} The calculated smear time is converted into phase bin units:
    $$ bins=\frac{\Delta t_{sec}}{Period_{sec}} \times n_{bins} $$

    \item \textbf{Kernel Construction:} A kernel is created as a 1D array of ones with a length equal to \texttt{smear$\_$bins}. This kernel is that the total power (sum of the kernel) remains unity. 

    \item \textbf{Convolution:} The pulse profile is convolved with this normalized boxcar kernel with \texttt{mode=`same'} parameter to ensure that the broadening is applied symmetrically, keeping the pulse centered while realistically degrading the resolution. 
\end{enumerate}

\subsection{Application}
This simulator also allows the analysis of the profile by switching on/off these ISM effects of scattering and DM smearing. These switches are passed as arguments in the \texttt{cli} module to compare how the profile appears when:
\begin{enumerate}[itemsep=0pt, parsep=0pt, topsep=1.0pt]
    \item Only scattering
    \item Only DM smearing
    \item Both scattering and DM smearing
    \item No broadening
\end{enumerate}
are distinctly applied. 

Thus, the observed pulse profile is:
\begin{equation}
    P(t)=P_{i}(t)*s(t)*D(t)*I(t)
\end{equation}
where, $P_{i}(t)$ is the intrinsic pulse shape, s(t) is the pulse scatter broadening due to propagation effects of ISM, D(t) is the dispersion smear across the narrow spectral channel and I(t) is the instrumental impulse response with * being the convolution function. 

%bandshape, circular convolution though fft of two functions, power law spectrum for timeseries and dmseries and make_dips (if not commented in the input file)
%scattering series - 4 models that can be implemented, dm smear
%make profile - broadening effects and make_fits - fits file 

\section{Noise Injection and Output Generation}
The final stage of the pipeline introduces observational artifacts to imitate real data. 
\subsection{System Noise and Bandshape}
\begin{enumerate}
    \item \textbf{Radiometer noise:} Random gaussian noise is added based on the target signal-to-noise ratio (snr).
    \item \textbf{Bandshape:} A \texttt{bandshape} utility simulates the non-flat frequency response of a receiver by convolving a square band with a Gaussian and adding low-level normal noise.
\end{enumerate}

\subsection{Interface with \texttt{psrchive}}
The \texttt{make$\_$fits} module creates a \texttt{psrchive.Archive} object. It populates the metadata (coordinates, telescope name, epoch, DM) and loads the synthesized data into the archive's sub-integrations. Finally, it calls the \texttt{unload()} method to save the simulation as a standard \textbf{PSRFITS} file, ready for timing analysis.

%% file: Chapters/chap5.tex
We have carried out the Gaussian fitting to get the profile parameters for all the 28 MSPs in the InPTA Cycle 45 data. A library of these parameters is made to then be used in the simulator to generate realistic simulations and is also a check of the accuracy of the model used and/or created by us. 

The primary check for the simulation pipeline is conducted through the \texttt{psrchive} software, utilizing its visualization and data reduction tools. By processing the simulated \textbf{PSRFITS} files, we can evaluate the accuracy of the synthetic profiles against observational expectations. 

\section{Integrated profile analysis}
The \textbf{PSRFITS} archives are frequency scrunched, that is, all the frequency channels are integrated together to get a profile with high signal-to-noise ratio. The standard \texttt{freq+} plot provides a multi-dimensional view of the pulsar signal. 

\begin{figure}[H] 
    \centering
    \includegraphics[width=0.7\linewidth]{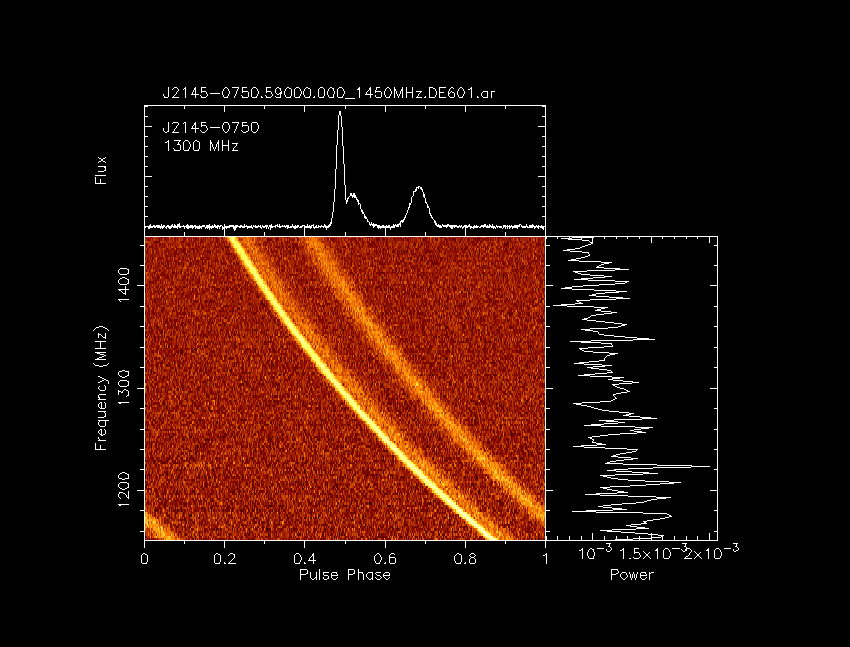}
    \caption[\texttt{freq+} plot for PSR J2145-0750.]{\footnotesize{\texttt{freq+} plot which gives flux as a function of phase (top panel), luminosity of the pulse across the frequency range as a function of phase (bottom panel) and frequency vs signal strength (right panel) for the simulated profile of PSR J2145-0750.}}
    \label{fig:5.1}
\end{figure}

\ref{fig:5.1} shows three critical panels to study the profile of the pulsar. The top panel (Flux vs Phase) shows the integrated profile after frequency scrunching. The profile of PSR J2145-0750 shows a complex structure with a primary peak and a distinct trailing component, reconstructed accurately from the multi-component Gaussian inputs and simulation. 

The bottom panel displays the pulse luminosity distributed across the observing band. It allows for the visual inspection of the signal's persistence across all frequency channels. This is for the case where the signal has not been de-dispersed and the time delay due to its DM across the frequency band is visible. To check that this is accurate, a test is performed which is discussed in the next section. 

The right panel shows the signal strength (power) as a function of frequency. It allows us to verify the application of the spectral index and the bandshape convolution applied to the profile (implemented in the \texttt{utils} module). 

\begin{figure}[H]
    \centering
    \includegraphics[width=0.7\linewidth]{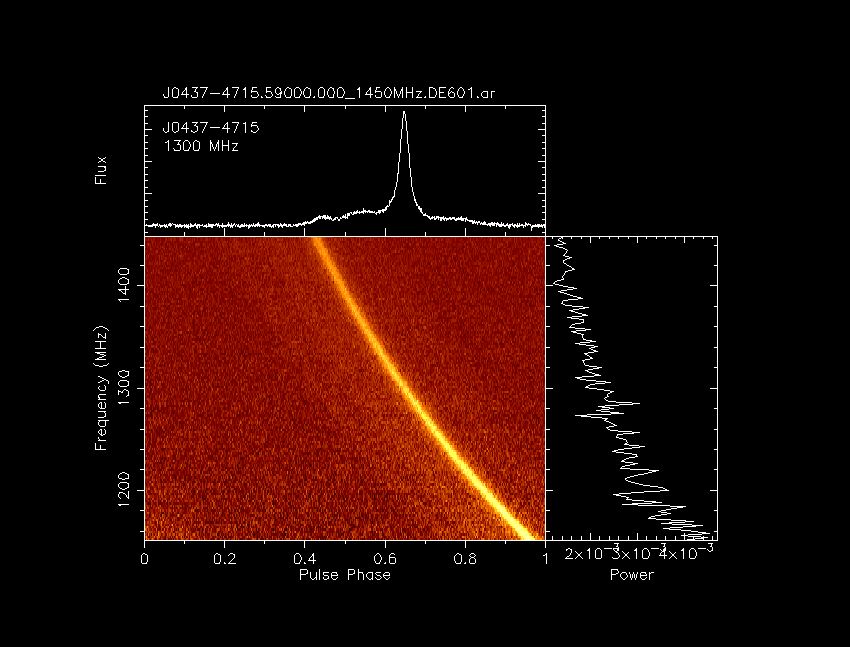}
    \caption[\texttt{freq+} plot for PSR J0437-4715.]{\footnotesize{\texttt{freq+} plot which gives flux as a function of phase (top panel), luminosity of the pulse across the frequency range as a function of phase (bottom panel) and frequency vs signal strength (right panel) for the simulated profile of PSR J0437-4715.}}
    \label{fig:5.2}
\end{figure}

\begin{figure}[H]
    \centering
    \includegraphics[width=0.7\linewidth]{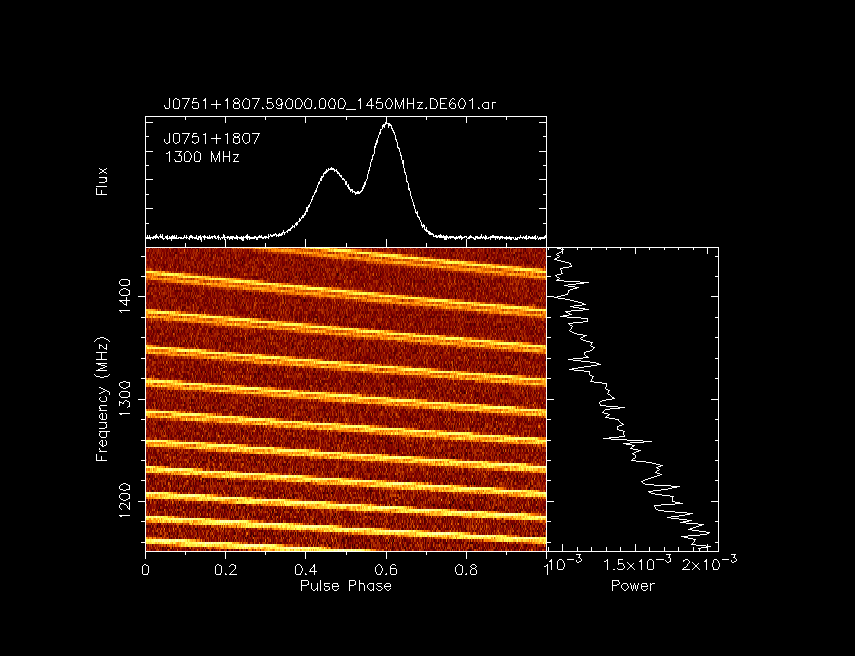}
    \caption[\texttt{freq+} plot for PSR J0751+1807.]{\footnotesize{\texttt{freq+} plot which gives flux as a function of phase (top panel), luminosity of the pulse across the frequency range as a function of phase (bottom panel) and frequency vs signal strength (right panel) for the simulated profile of PSR J0751+1807.}}
    \label{fig:5.3}
\end{figure}

\section{Impact of ISM}
\subsection{Effects of DM delay without dedispersion}
When the simulated data is viewed without dedispersion, as is seen in the bottom panel of the figure \ref{fig:5.1} and \ref{fig:5.2}, the characteristic quadratic sweep of the signal across the frequency-phase plane becomes evident. The pulse arrives earlier at higher frequencies than at the lower frequencies. This delay is directly due to : 
\begin{equation}
    \Delta t = 4.15 \times 10^{6} \text{ ms} \times DM \times (\nu_{low}^{-2}-\nu_{high}^{-2})
\end{equation}
where $\nu_{low}$ and $\nu_{high}$ are the frequencies in GHz. The simulator correctly implements this time delay by populating the sub-integrations according to the frequency-dependent arrival times. This is checked by calculating the $\Delta t$ for these pulsars by taking the DM from \hyperlink{https://www.atnf.csiro.au/research/pulsar/psrcat/}{ATNF Pulsar Catalogue} and $\nu_{low}$ and $\nu_{high}$ as 1.15 GHz and 1.45 GHz respectively, as the frequency range in which the simulation was carried out. The luminous regions of the f vs phase plots is the indicator of the pulse. The number of those regions would provide us with the number of periods passed due to the time delay. Therefore, time delay divided by the time period of that pulsar gives the delay as multiples of the time period which should be equal to the number of bright regions (periods) of the pulsar. The time period p0 of the pulsar is also obtained from the \hyperlink{https://www.atnf.csiro.au/research/pulsar/psrcat/}{ATNF Pulsar Catalogue}

For example, for PSR J0740+6620 in the range of 1.15 to 1.45 GHz with DM=14.96:
$$ \Delta t = 4.15 \times 10^{6} \times 14.96 \times (1.15^{-2}-1.45^{-2}) \text{ ms} = 17.42 \text{ ms} $$
$$ \Delta t/p0 = 17.42/2.8 = 6.22 $$
\begin{figure}[H]
    \centering
    \includegraphics[width=0.6\linewidth]{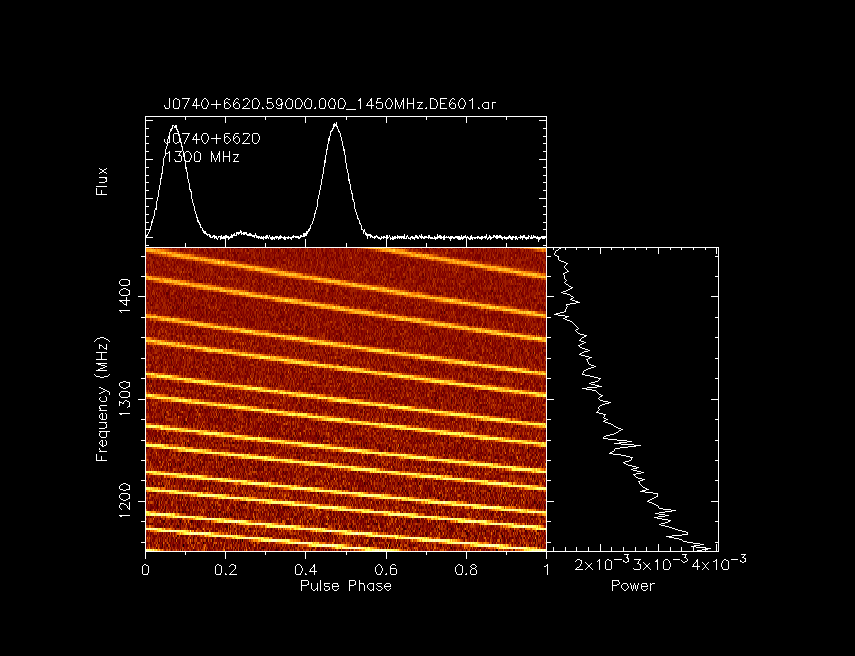}
    \caption[\texttt{freq+} plot for PSR J0740+6620.]{\footnotesize{\texttt{freq+} plot which gives flux as a function of phase (top panel), luminosity of the pulse across the frequency range as a function of phase (bottom panel) and frequency vs signal strength (right panel) for the simulated profile of PSR J0740+6620.}}
    \label{fig:5.4}
\end{figure}
which gives the number of periods matching the simulation that is 6. Similarly, it can be seen for PSR J0751+1807 in \ref{fig:5.3}, PSR J2145-0750 validating the correct implementation of the DM delay in the simulation. 
\begin{enumerate}
    \item \textbf{PSR J0751+1807: } DM = 30.24, p0 = 3.48 ms
    $$ \Delta t = 4.15 \times 10^{6} \times 30.24 \times (1.15^{-2}-1.45^{-2}) \text{ ms} = 35.20 \text{ ms} $$
    $$ \Delta t/p0 = 35.20/3.48 = 10.11 $$

    \item \textbf{PSR J2145-0750: } DM = 9.01, p0 = 16.05 ms
    $$ \Delta t = 4.15 \times 10^{6} \times 9.01 \times (1.15^{-2}-1.45^{-2}) \text{ ms} = 0.65 \text{ ms} $$
    $$ \Delta t < p0 \implies 1 \text{ period} $$
\end{enumerate}

\subsection{Scatter Broadening and DM smear}
Scatter broadening is evident with tail broadening of the profile at low frequencies as $\tau_{sc} \propto \nu^{\alpha}$ where $\alpha$ is usually -4.4. By switching off DM smear and simulating in the frequency range of 300 to 500 MHz, only the effect of scattering is apparent:   
\begin{figure}[H]
\centering
\begin{subfigure}{.5\textwidth}
  \centering
  \includegraphics[width=.9\linewidth]{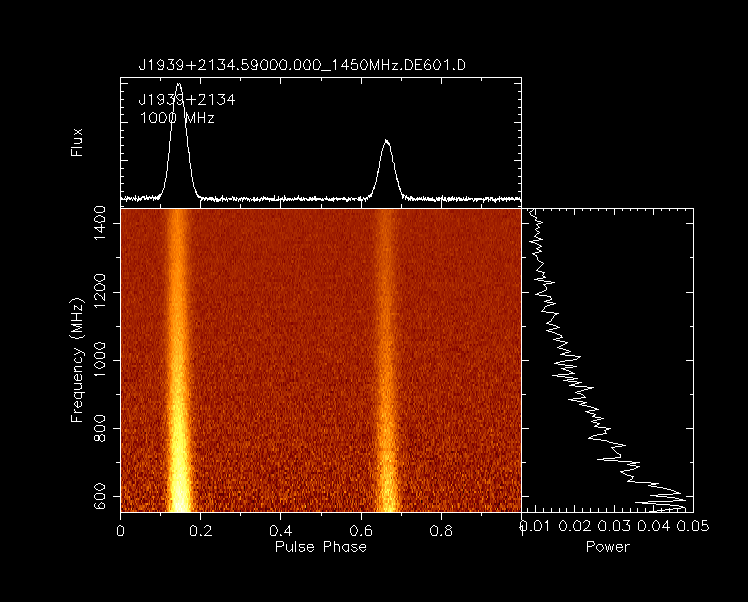}
  \caption{\footnotesize {Intrinsic simulated profile with no scattering and no DM smearing}}
  \label{fig:5.5(a)}
\end{subfigure}%
\begin{subfigure}{.5\textwidth}
  \centering
  \includegraphics[width=.9\linewidth]{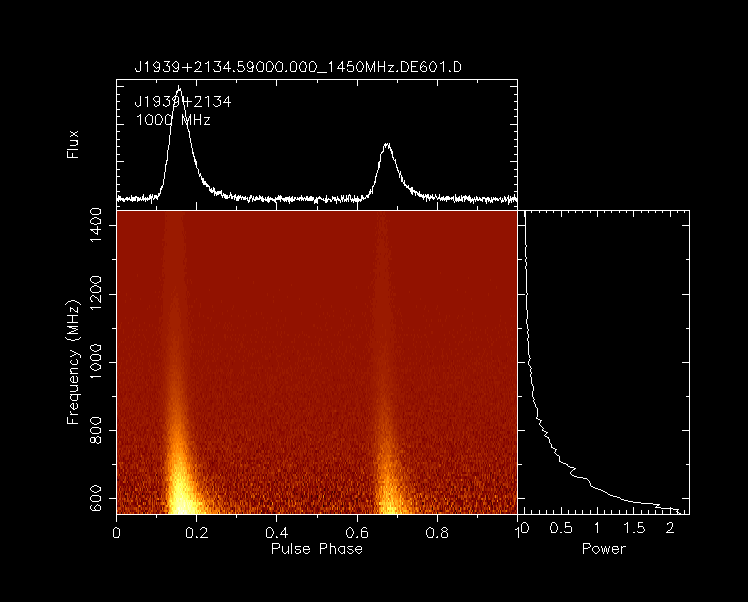}
  \caption{\footnotesize{Simulated profile with only scattering effects and no DM smearing}}
  \label{fig:5.5(b)}
\end{subfigure}
\caption{\footnotesize {Simulated profiles of PSR J1939+2134 in the frequency range of 550 to 1450 MHz to understand the effect of scattering. The tail broadening in (b) is due to the time delay in the arrival of pulses because of scattering}}
\label{fig:5.5}
\end{figure}

Subsequently, switching off scattering and just viewing the effect of smearing ($\Delta \tau_{DM} \propto \nu^{-3}\times DM$), gives the symmetrical increase in width of the profile. It is more prominent when taken over a larger frequency range. A slight tilting on the left side of the bright pulse region at lower frequencies is observed due to the smearing effect. 
\begin{figure}[H]
\centering
\begin{subfigure}{.5\textwidth}
  \centering
  \includegraphics[width=.9\linewidth]{Chapters/nodm_71.png}
  \caption{\footnotesize{Intrinsic profile with no scattering and no DM smearing}}
  \label{fig:5.6(a)}
\end{subfigure}%
\begin{subfigure}{.5\textwidth}
  \centering
  \includegraphics[width=.9\linewidth]{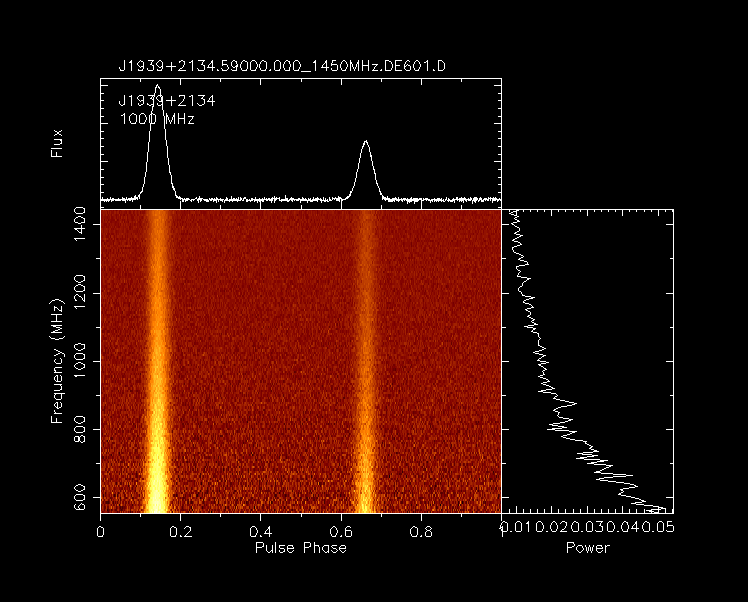}
  \caption{\footnotesize{Profile with only DM smearing, and no scattering effect}}
  \label{fig:5.6(b)}
\end{subfigure}
\caption{\footnotesize{Simulated profiles of PSR J1939+2134 in the frequency range of 550 to 1450 MHz to understand the effect of DM smearing. The slight tilting of the luminous regions in (b) is due to the smearing time delay at lower frequencies.}}
\label{fig:5.6}
\end{figure}

\section{Residual Analysis}
\subsection{Data and simulation comparison}
To quantify the success of the simulation, we perform a Residual Analysis ($\epsilon_{i}=y_{i}-f{i}$). Physically, they represent the portion of the data that the model fails to explain. For the comparison, the phase bins of the simulated profiles are made equal to the observed data (typically nbin=1024) and aligned precisely using a circular roll to ensure maximum overlap of the primary peaks. 

In a successful model, residuals should behave as independent random samples from a probability distribution (typically a normal distribution) with a mean of zero. A plot of residuals against the independent variable is physically important because it reveals systematic deviations. If the residuals show a visible pattern or trend rather than random noise, it indicates that the functional form of the model is incorrect or that an external physical factor has been neglected. Residual analysis helps confirm that the experimental uncertainties are indeed random fluctuations, such as thermal noise, rather than avoidable mistakes or systematic instrumental bias \citep{berendsen2011data}. 

As illustrated, the residual panel (bottom) shows random noise with no significant remaining structural features. This indicates that the multi-component Gaussian model effectively captures the physical emission.

Of the 28 MSPs, there were 10 objects whose Band 5 data were neither available in the InPTA dataset nor in the EPN database. For those sources, InPTA Band 3 data were utilized. PSR J1705-1903 had a very low signal to noise ratio of 5 and hence has not been used. All pulse profiles are baseline-subtracted and linearly scaled for comparison; the term ‘normalized intensity’ is used in this relative sense rather than strict peak or flux normalization.

\begin{enumerate}
    \item \textbf{Band 3 results: }
    \begin{figure}[H]
        \centering
        \begin{subfigure}{0.55\textwidth}
        \centering
            \includegraphics[width=0.8\linewidth]{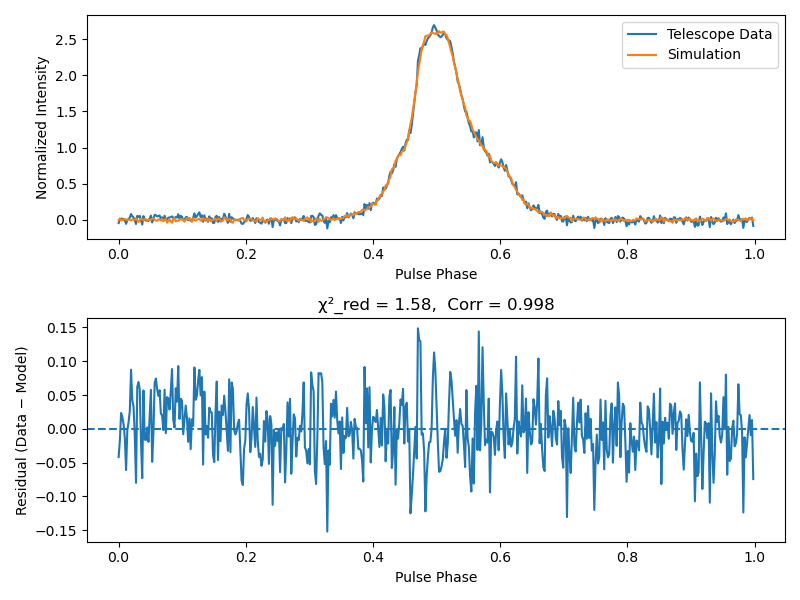}
            \caption{PSR J1125+7819}
            \label{fig:5.8(a)}
        \end{subfigure}%
        \begin{subfigure}{0.55\textwidth}
            \centering
            \includegraphics[width=0.8\linewidth]{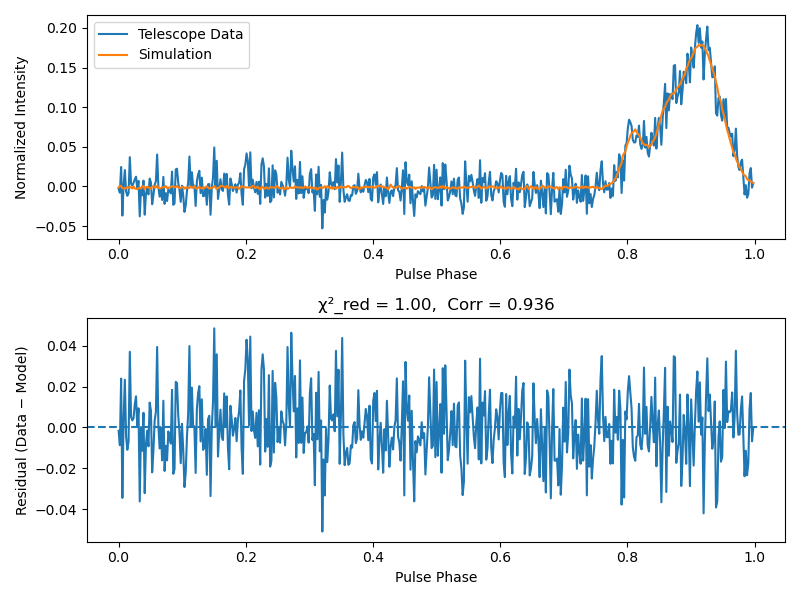}
            \caption{PSR J0614-3329}
            \label{fig:5.8(b)}
        \end{subfigure}
    \end{figure}    

    \begin{figure}[H]
        \ContinuedFloat
        \begin{subfigure}{0.55\textwidth}
            \centering
            \includegraphics[width=0.9\linewidth]{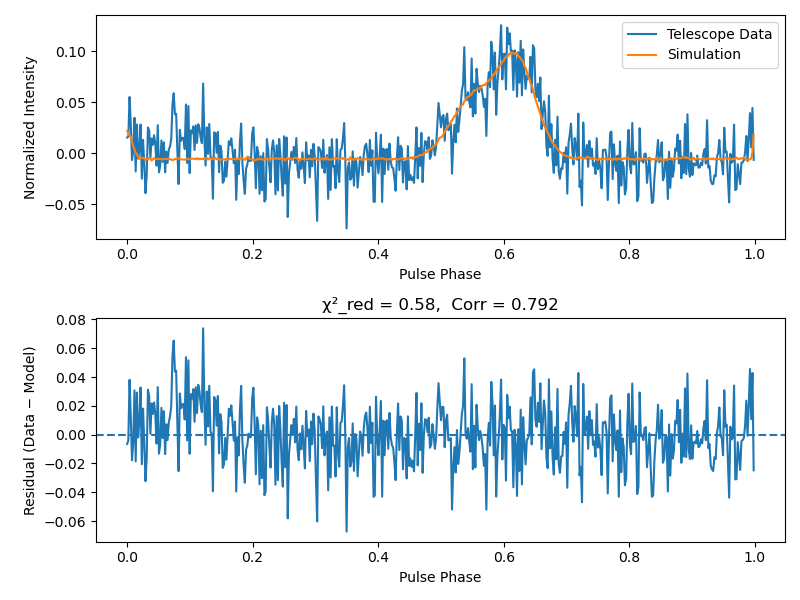}
            \caption{PSR 0030+0451}
            \label{fig:5.8(c)}
        \end{subfigure}%
        \begin{subfigure}{0.55\textwidth}
            \centering
            \includegraphics[width=0.9\linewidth]{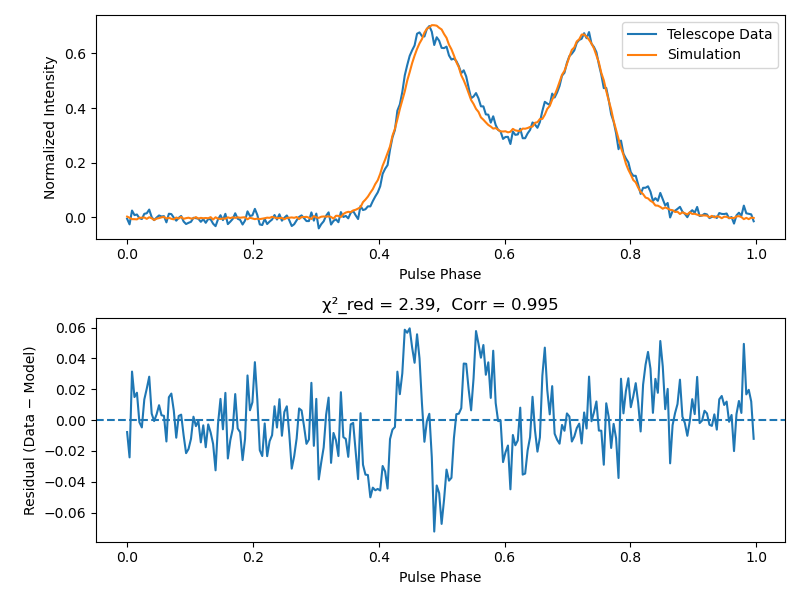}
            \caption{PSR 0034-0534}
            \label{fig:5.8(d)}
        \end{subfigure}
        \vfill
        \begin{subfigure}{0.55\textwidth}
            \centering
            \includegraphics[width=0.9\linewidth]{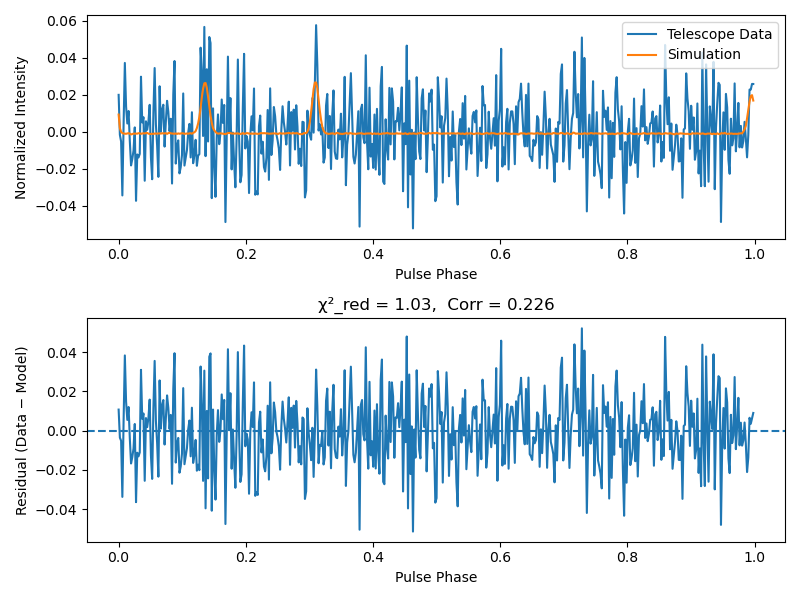}
            \caption{PSR 0610-2100}
            \label{fig:5.8(e)}
        \end{subfigure}%
        \begin{subfigure}{0.55\textwidth}
            \centering
            \includegraphics[width=0.9\linewidth]{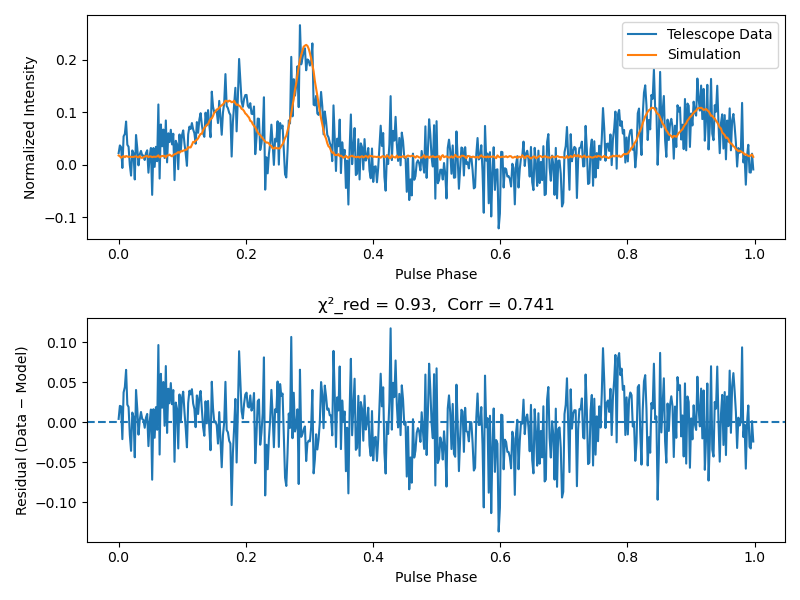}
            \caption{PSR 1545-4550}
            \label{fig:5.8(f)}
        \end{subfigure}
        \vfill
        \begin{subfigure}{0.55\textwidth}
            \centering
            \includegraphics[width=0.9\linewidth]{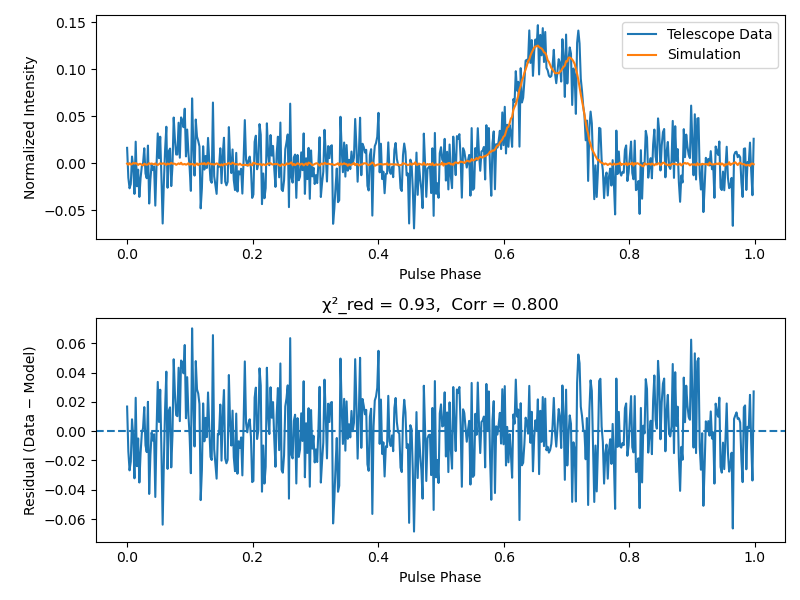}
            \caption{PSR J1745+1017}
            \label{fig:5.8(g)}
        \end{subfigure}%
        \begin{subfigure}{0.55\textwidth}
            \centering
            \includegraphics[width=0.9\linewidth]{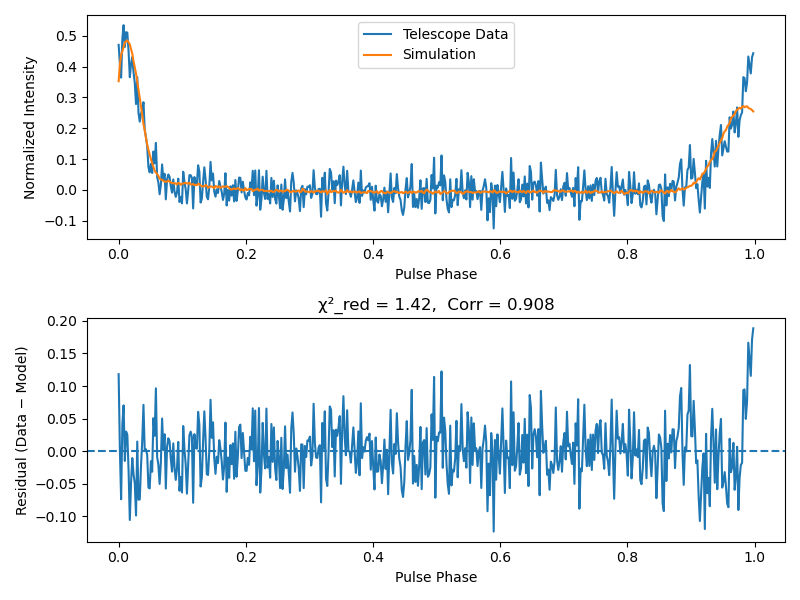}
            \caption{PSR 1910+1256}
            \label{fig:5.8(h)}
        \end{subfigure}
        \vfill
        \centering
        \begin{subfigure}{0.6\textwidth}
            \centering
            \includegraphics[width=0.8\linewidth]{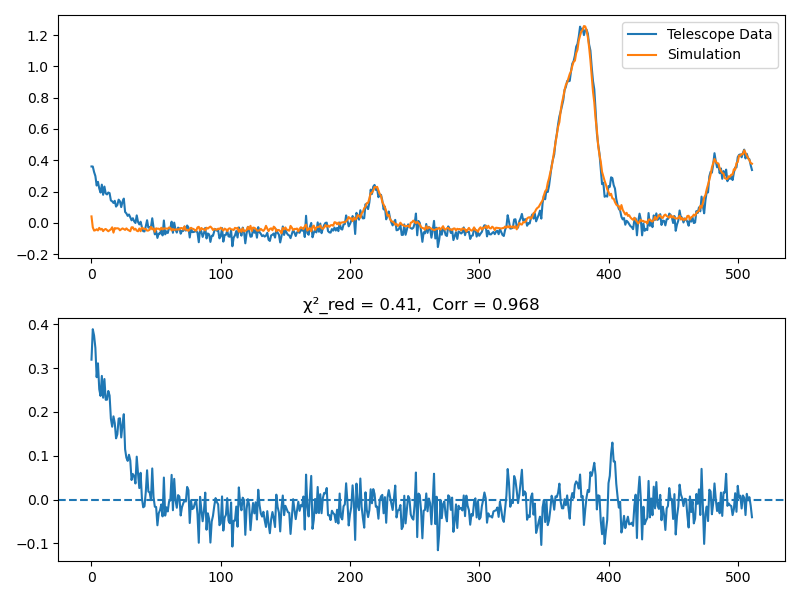}
            \caption{PSR J2302+4442}
            \label{fig:5.8(i)}
        \end{subfigure}
        \label{fig:5.8}
        \caption{Comparison between data and model for Band 3 data}
    \label{5.7}
    \end{figure}

    \newpage
    \item \textbf{Band 5 results: } 
    \begin{figure}[H]
    \centering
    \begin{subfigure}{.55\textwidth}
        \centering
        \includegraphics[width=.9\linewidth]{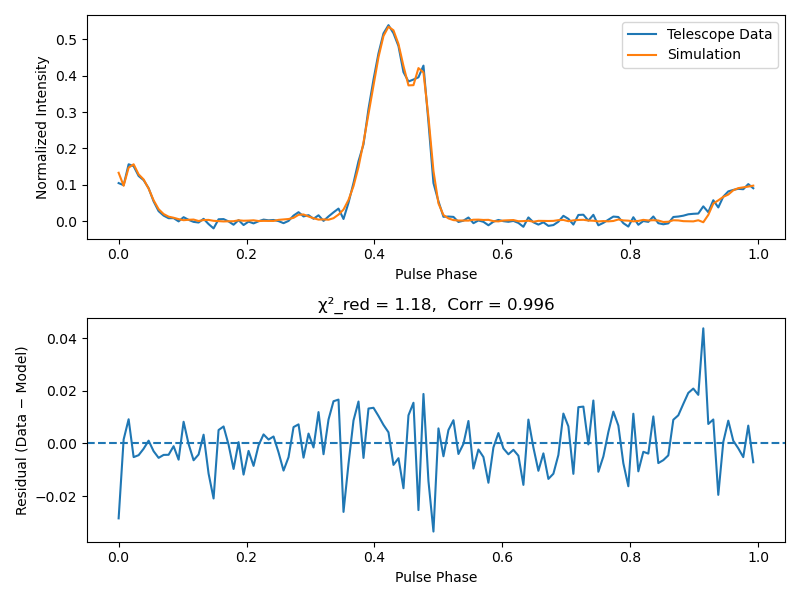}
        \caption{PSR J1857+0943}
        \label{fig:5.7(a)}
    \end{subfigure}%
    \begin{subfigure}{.55\textwidth}
        \centering
        \includegraphics[width=.9\linewidth]{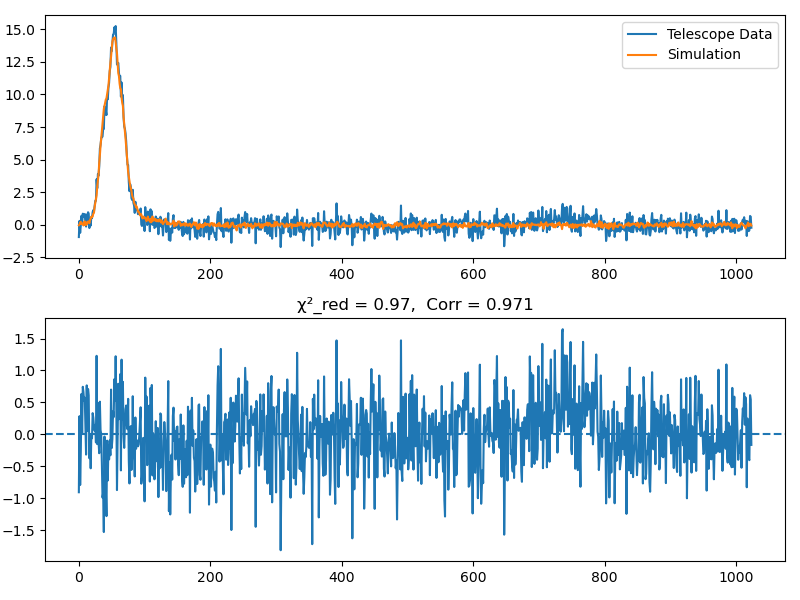}
        \caption{PSR J1744-1134}
        \label{fig:5.7(b)}
    \end{subfigure}
    \vfill
    \begin{subfigure}{.55\textwidth}
        \centering
        \includegraphics[width=0.9\linewidth]{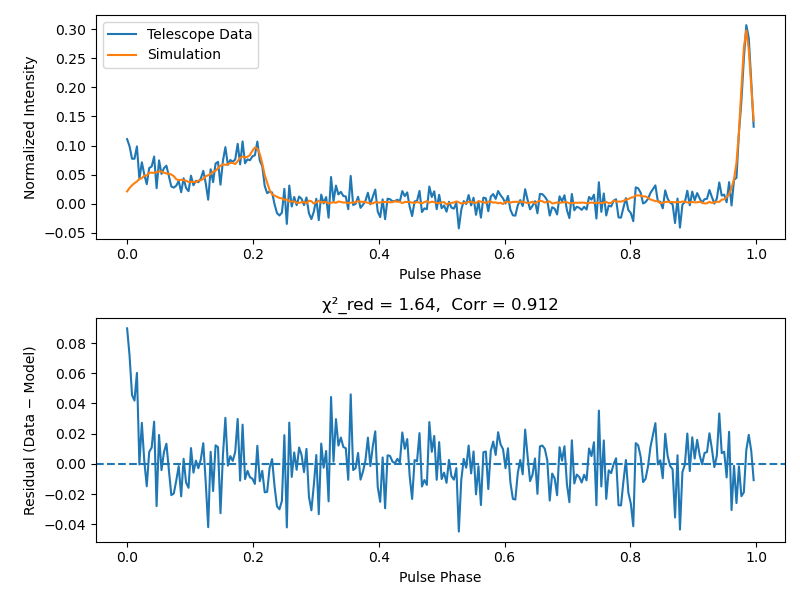}
        \caption{PSR J2145-0750}
        \label{fig:5.7(c)}
    \end{subfigure}%
    \begin{subfigure}{0.55\textwidth}
        \centering 
        \includegraphics[width=0.9\linewidth]{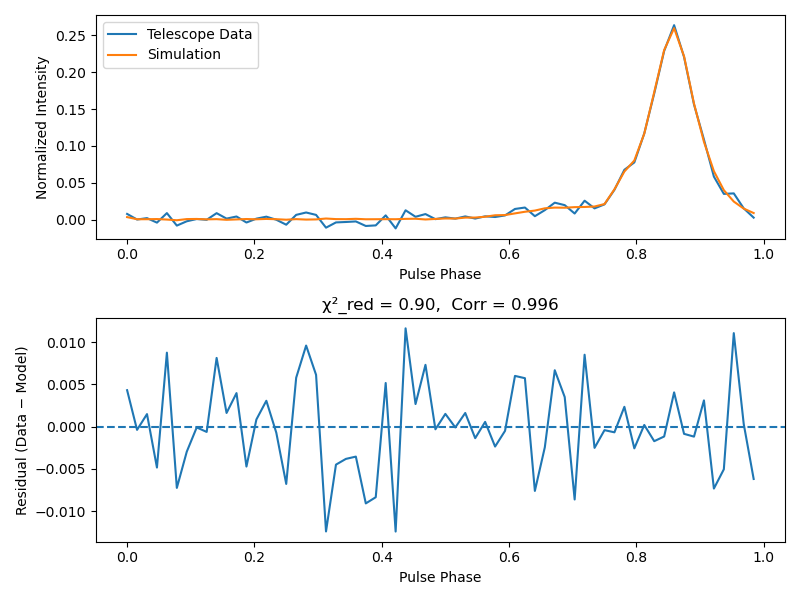}
        \caption{PSR J1643-1224}
        \label{fig:5.7(d)}
    \end{subfigure}
    \vfill
    \begin{subfigure}{0.55\textwidth}
        \centering
        \includegraphics[width=0.9\linewidth]{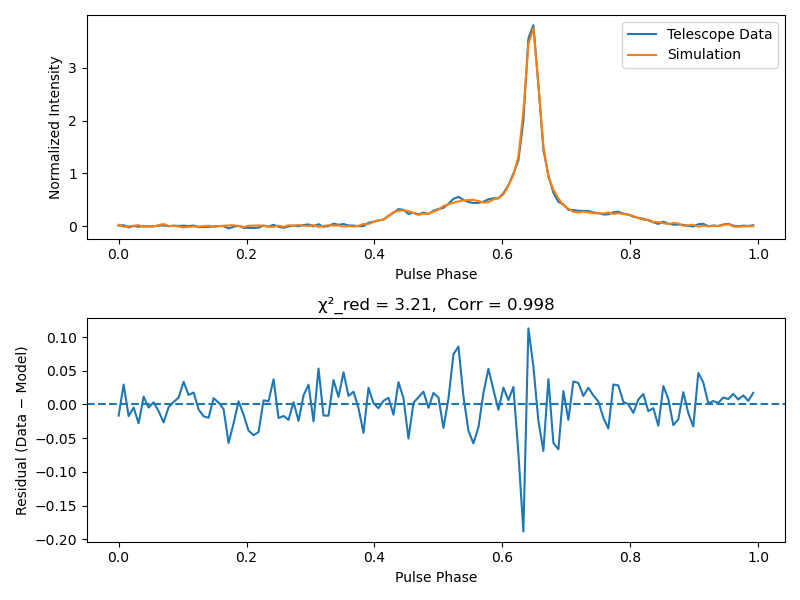}
        \caption{PSR J0437-4715}
        \label{fig:5.7(e)}
    \end{subfigure}%
    \begin{subfigure}{0.55\textwidth}
        \centering
        \includegraphics[width=0.9\linewidth]{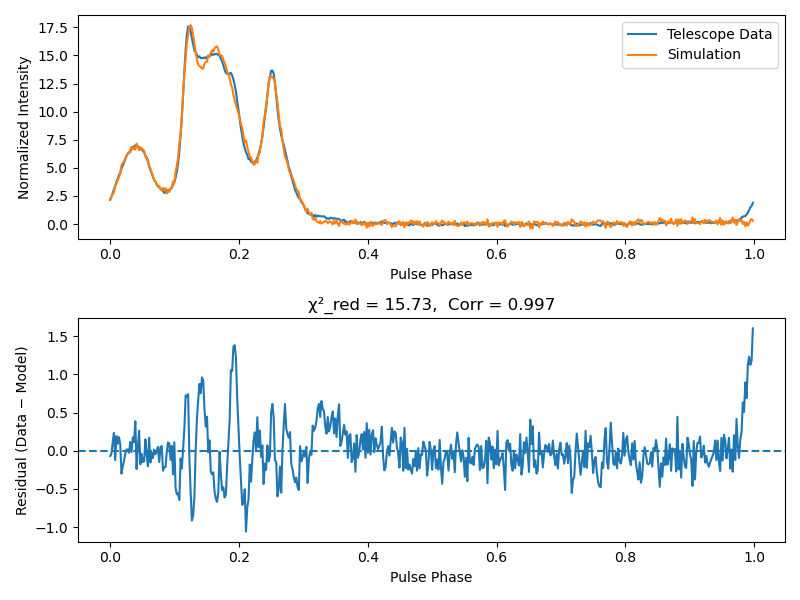}
        \caption{PSR J0613-0200}
        \label{fig:5.7(f)}
    \end{subfigure}
    \end{figure}

    \begin{figure}[H]
        \ContinuedFloat
        \centering
    \begin{subfigure}{.55\textwidth}
        \centering
        \includegraphics[width=.9\linewidth]{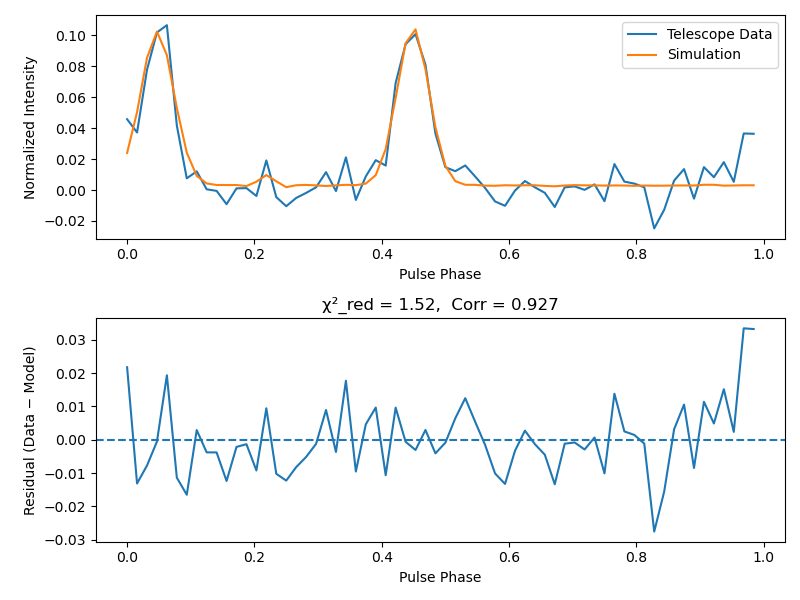}
        \caption{PSR J0740+6620}
        \label{fig:5.7(g)}
    \end{subfigure}%
    \begin{subfigure}{.55\textwidth}
        \centering
        \includegraphics[width=.9\linewidth]{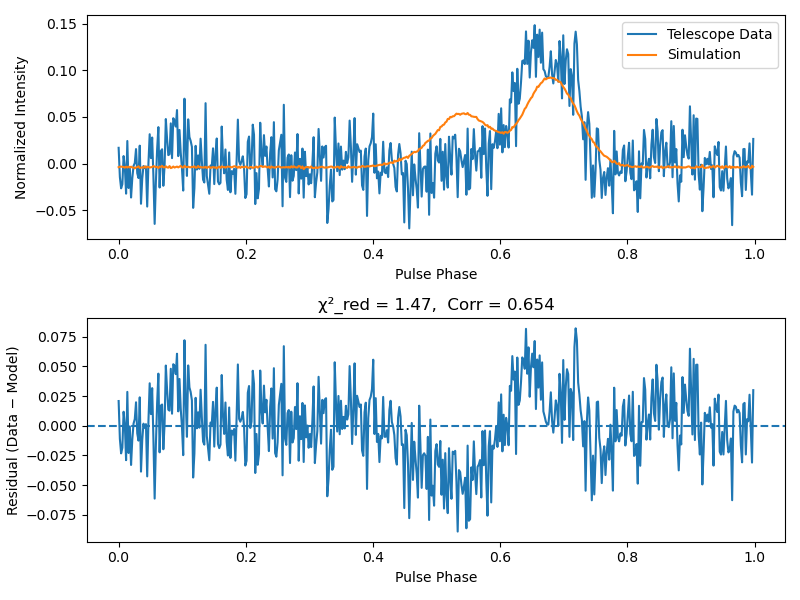}
        \caption{PSR J0752+1807}
        \label{fig:5.7(h)}
    \end{subfigure}
    \vfill
    \begin{subfigure}{.55\textwidth}
        \centering
        \includegraphics[width=0.9\linewidth]{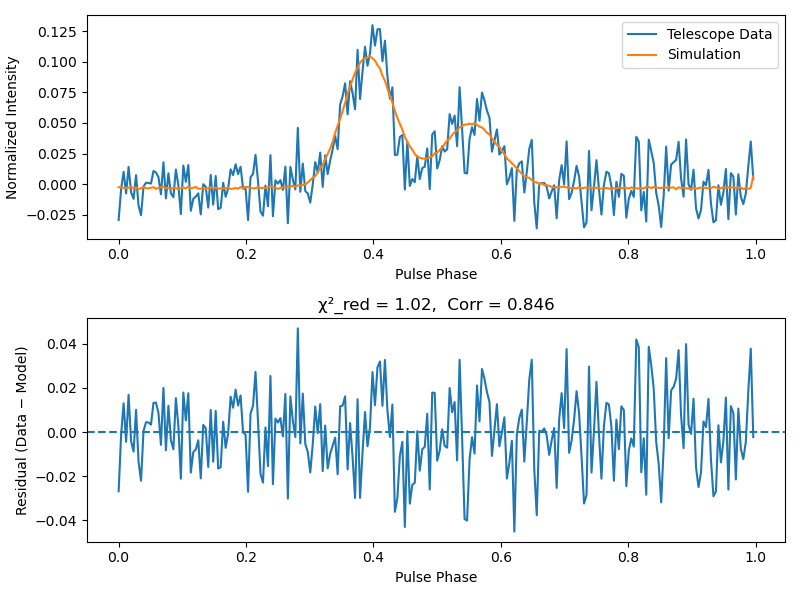}
        \caption{PSR J0900-3144}
        \label{fig:5.7(i)}
    \end{subfigure}%
    \begin{subfigure}{0.55\textwidth}
        \centering 
        \includegraphics[width=0.9\linewidth]{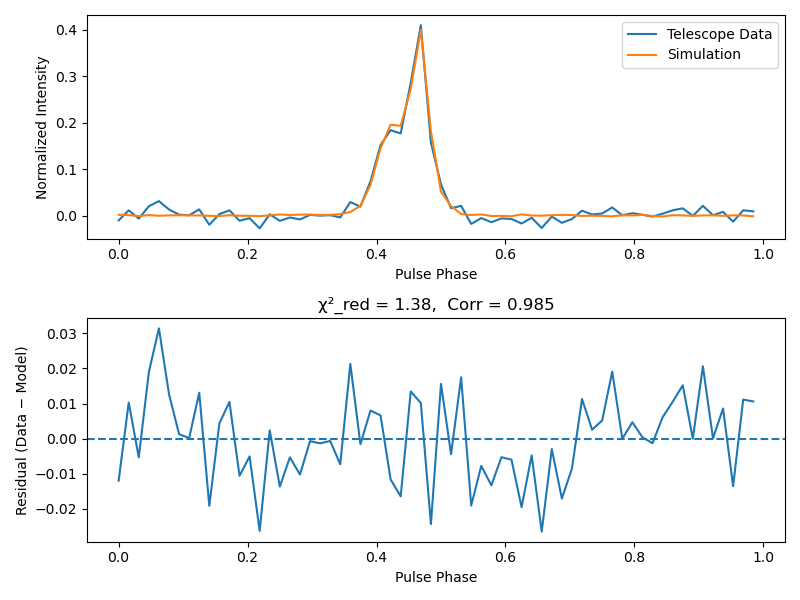}
        \caption{PSR J1600-3053}
        \label{fig:5.7(j)}
    \end{subfigure}
    \vfill
    \begin{subfigure}{0.55\textwidth}
        \centering
        \includegraphics[width=0.9\linewidth]{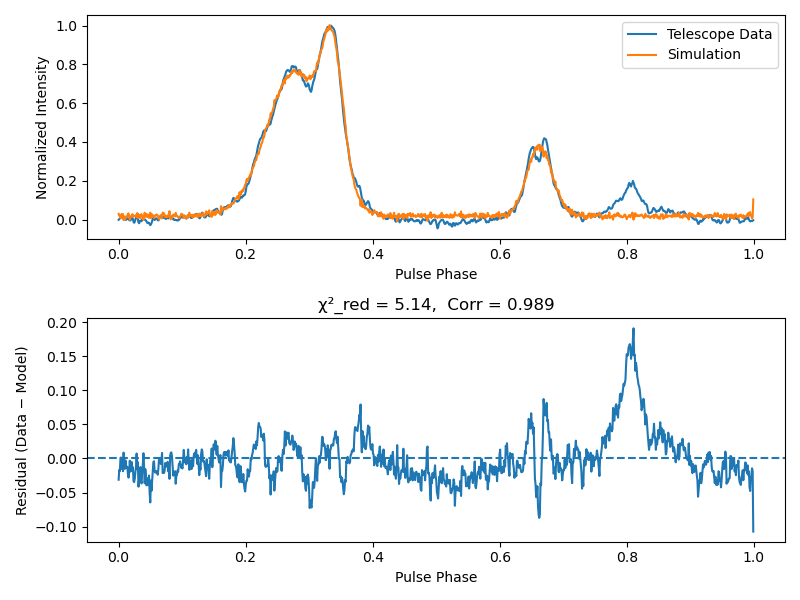}
        \caption{PSR J1012+5307}
        \label{fig:5.7(k)}
    \end{subfigure}%
    \begin{subfigure}{0.55\textwidth}
        \centering
        \includegraphics[width=0.9\linewidth]{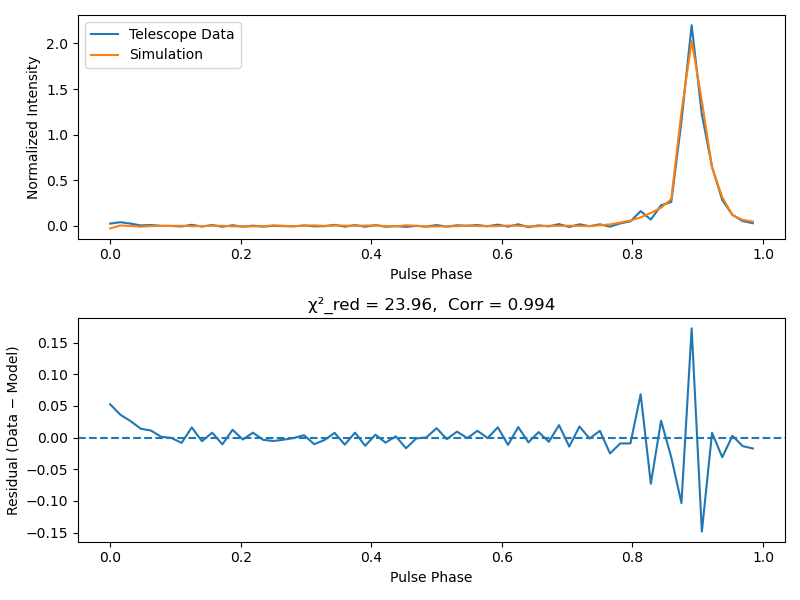}
        \caption{PSR J1713+0740}
        \label{fig:5.7(l)}
    \end{subfigure}
    \vfill
    \begin{subfigure}{.55\textwidth}
        \centering
        \includegraphics[width=.9\linewidth]{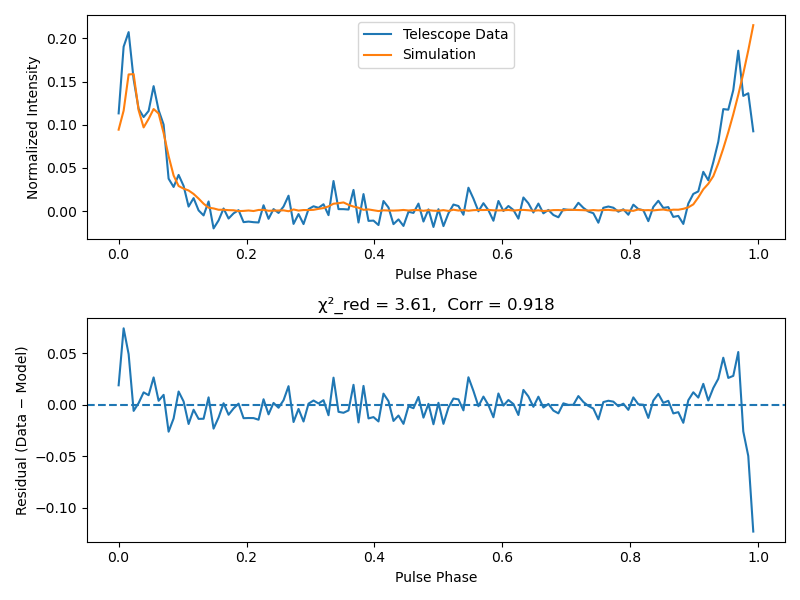}
        \caption{PSR J1730-2304}
        \label{fig:5.7(m}
    \end{subfigure}%
    \begin{subfigure}{.55\textwidth}
        \centering
        \includegraphics[width=.9\linewidth]{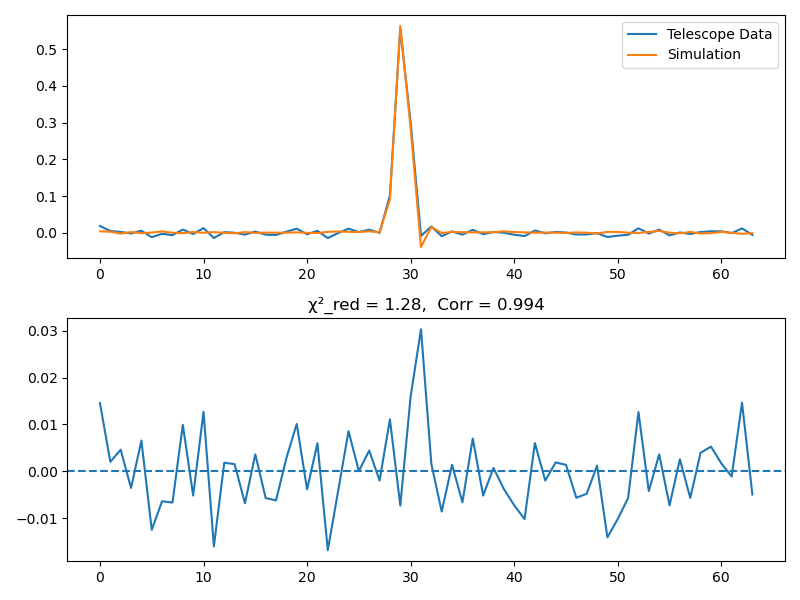}
        \caption{PSR J1909-3744}
        \label{fig:5.7(n)}
    \end{subfigure}
    \end{figure}
    \begin{figure}[H]
        \ContinuedFloat
        \centering
    \begin{subfigure}{.55\textwidth}
        \centering
        \includegraphics[width=0.9\linewidth]{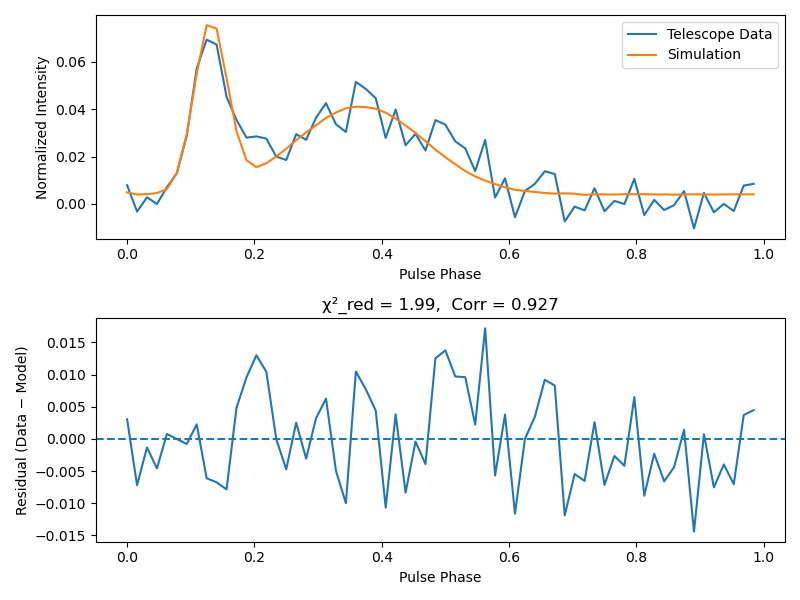}
        \caption{PSR J1944+0907}
        \label{fig:5.7(o)}
    \end{subfigure}%
    \begin{subfigure}{0.55\textwidth}
        \centering 
        \includegraphics[width=0.9\linewidth]{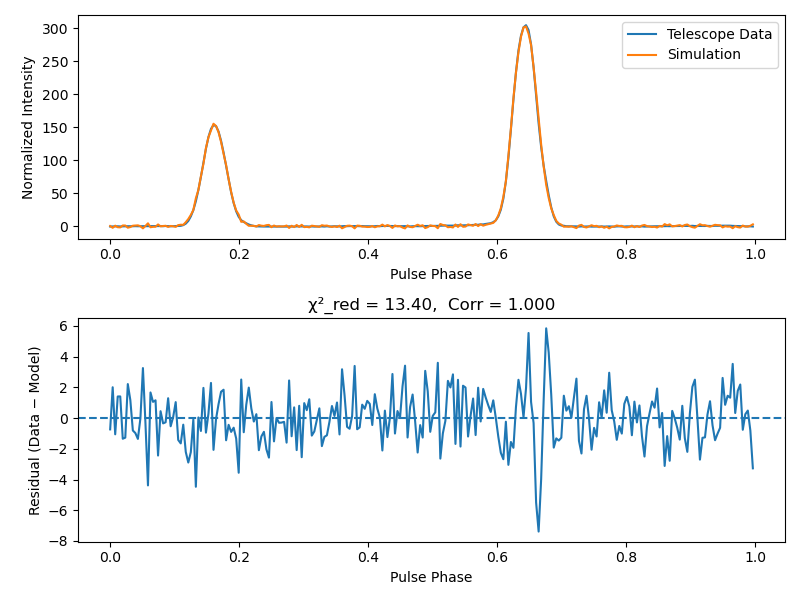}
        \caption{PSR J1939+2134}
        \label{fig:5.7(p)}
    \end{subfigure}
    \vfill
    \begin{subfigure}{0.55\textwidth}
        \centering
        \includegraphics[width=0.9\linewidth]{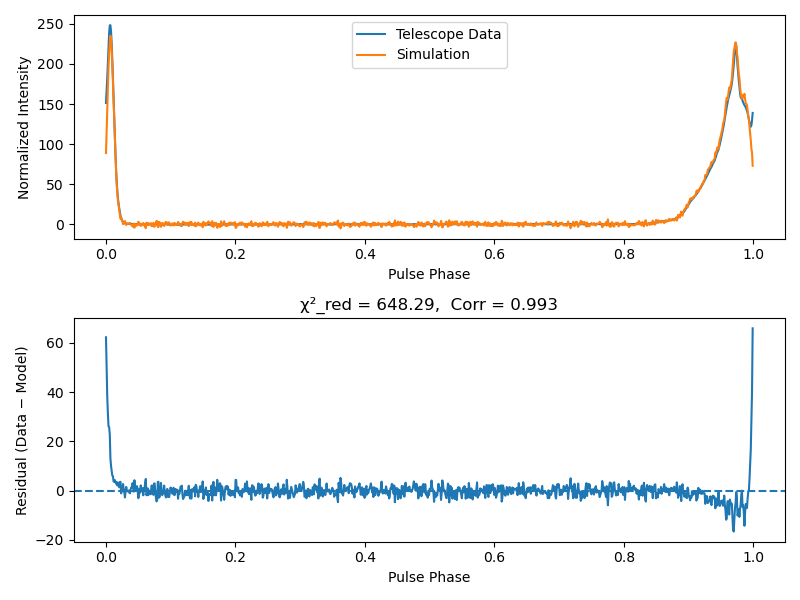}
        \caption{PSR J1022+1001}
        \label{fig:5.7(q)}
    \end{subfigure}%
    \begin{subfigure}{0.55\textwidth}
        \centering
        \includegraphics[width=0.9\linewidth]{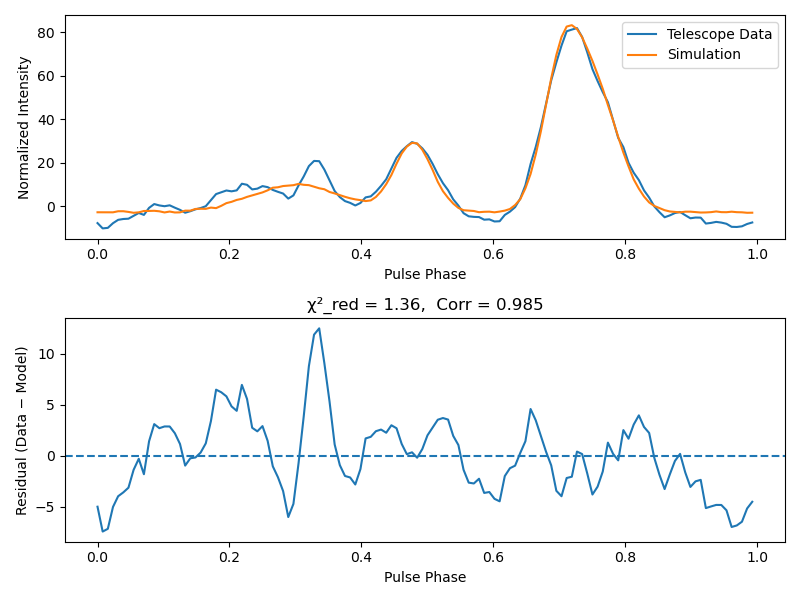}
        \caption{PSR J2124-3358}
        \label{fig:5.7(r)}
    \end{subfigure}
    \caption{Comparison between data and model for Band 5 data}
    \label{fig:5.8}
    \end{figure}

\end{enumerate}

%attach bad results as well indicating there may be some artifacts missing in the simulation - not entirely perfect 

\subsection{Residual width}
It is known through years of observation of millisecond pulsars and normal pulsars that the latter evolves with frequency in terms of their number of profile components and intrinsic widths. Whereas, widths of MSP profiles are essentially independent of frequency suggests that the emission in these pulsars originates from the same height at multiple frequencies \citep{LorimerKramer2004, Kramer1999MSP}. This is expected as the radius of the light cylinder and, hence, the size of the magnetosphere is much smaller for MSPs. This does not leave much space for the active emission to spread out. 
Comparison of profiles by simulation in bands 3 and 5, no major changes in the component widths and separation were observed. The effect was only due to the ISM and no major intrinsic frequency evolution can be concluded. 

To further confirm the observation, the residual widths $$ res_{width} = wd - wm $$ where wd is the component width of the profile from the data and wm is that of the simulated profile. These were calculated separately for each bands 5 and 3 and were compared by combining them on a single plot as a function of frequency. 

\begin{figure}[H]
    \centering
    \includegraphics[width=0.8\linewidth]{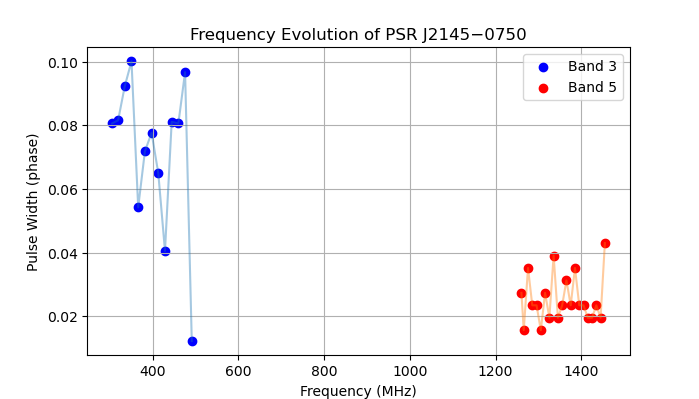}
    \caption{Residual width (component width of the simulated profile - component width of the data) of PSR J2145-0750 in Band 3 and Band 5}
    \label{fig:5.9}
\end{figure}

\begin{figure}[H]
    \centering
    \includegraphics[width=0.8\linewidth]{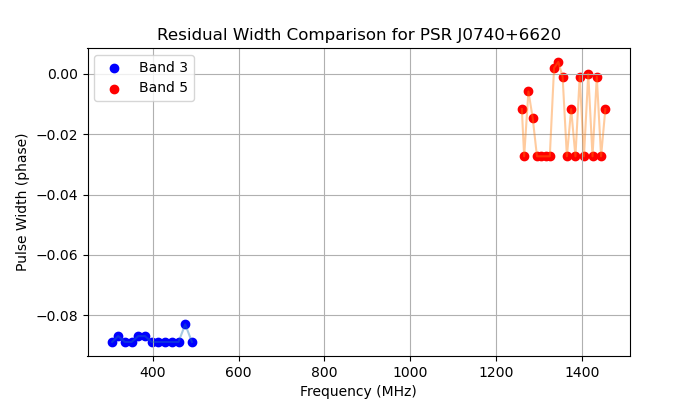}
    \caption{Residual width of PSR J0740+6620 in Band 3 and Band 5}
    \label{fig:5.11}
\end{figure}

\begin{figure}[H]
    \centering
    \includegraphics[width=0.8\linewidth]{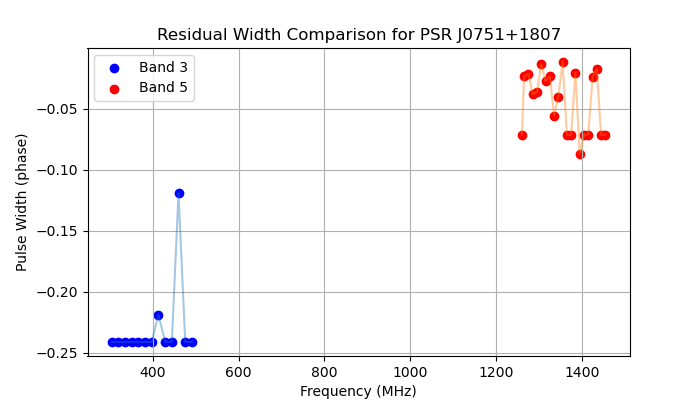}
    \caption{Residual width of PSR J0751+1807 in Band 3 and Band 5}
    \label{fig:5.12}
\end{figure}

But for PSR J2145-0750 (Figure \ref{fig:5.9}), the residual of the width in band 3 (low frequency) is more than the residual of the width in band 5 (higher frequency) which provides us with the evidence of frequency evolution of the widths of the components. The component width of the data cannot be more than the component width of the model (simulated profile), as all the effects of ISM have been taken into account that can increase the width (scattering scaled as $\nu^{-4.4}$ and DM smear). On the other hand, for other pulsars such as PSR J0740+6620, PSR J0751+1807 in Figures \ref{fig:5.11} and \ref{fig:5.12}, it is the other way around, where the residual width of band 3 is smaller and is, in fact, negative. This shows that the component width of the data is less than that of the model in band 3 implying that the effect of ISM is shallower than what is utilized in the model. 

Since all the other pulsars show the same trend, it proves that the anomaly in PSR J2145-0750 is not due to the external effects as all of them have been taken care of, but due to the intrinsic frequency evolution. As in the simulation input file, the only parameters not included are the width, height and phase scaling with frequency. This correctly matches with the already existing literature \cite{Kramer1999MSP}. 

\begin{figure}[H]
    \centering
    \includegraphics[width=0.8\linewidth]{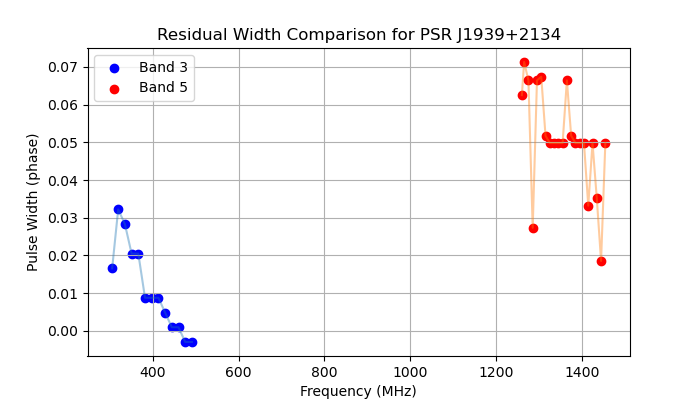}
    \caption{Residual width of PSR J1939+2134 in Band 3 and Band 5}
    \label{fig:5.13}
\end{figure}

The model is user-driven, and the propagation parameters can be varied to make the simulations more in accordance with the data. This is the utility of this software that allows us to model the ISM effects accurately along with other properties of the pulsar itself. 

\subsection{\texorpdfstring{$\chi_{red}^{2}$}{chi	extsubscript{red}	extsuperscript{2}} Analysis}
After checking the randomness of the residuals and accepting the functional choices of the model, we can proceed to the next step, that is, the chi-squared test. $\chi^{2}$ is the weighted sum of squared deviations, where each term is normalized by the variance ($\sigma^{2}$), $\sigma$ obtained in this work as the mean absolute deviation (MAD) of the off-pulse region: 
$$ \chi^{2} = \sum_{i=1}^{n} \frac{(y_{i}-f_{i})^{2}}{\sigma_{i}^{2}} $$
The value of reduced chi square, $\chi^{2}_{red}$ is this sum divided by the number of degrees of freedom ($\nu=n-m$), where n is the number of data points and m is the number of parameters. $\chi^{2}_{red}\sim 1$ indicates that the residuals are entirely consistent with the estimated noise level of the experiment. It suggests that the model is a statistically justified representation of the physical process. Higher value $\chi^{2}_{red}>> 1$ (under-fitting) suggests that the data deviate significantly from the function, meaning the model is insufficient to describe the physical reality or that the experimental errors were underestimated. Whereas $\chi^{2}_{red} << 1$, value much lower than 1 (over-fitting), means that the model has too many parameters and is fitting the noise, rather than the underlying physics. 
%how and why reduced chi square being near to 1 or more than tells us about the profile's properties - how something is maybe missing or everything is being captured in a few pulsars with examples. 

\subsection{Limitations}
The primary limitation of this thesis is that the effect of scintillation has not been included in the model. This work was carried out first only for the sources utilized in pulsar timing by the InPTA. This can be further extended to include other pulsars (MSP and canonical both) with specific profile evolution parameters scaling with frequency. Although gaussian fitting gives a good approximation for the profiles, other methods like Lorentzian, Profile Component Analysis (PCA), where profile can be broken into multiple basis function can be used to give better approximations for the parameters.

\section{Conclusion}
This simulator is highly useful for Pulsar Timing Array (PTA) research. By producing realistic \textbf{PSRFITS} files that include stochastic DM variations and discrete ISM events, researchers can test the robustness of timing pipelines against simulated `noise' and ISM-induced delays. The ability to simulate stable templates with known parameters allows for the precise measurement of Time of Arrivals (TOAs) and the evaluation of jitter noise.

Out of the 9 pulsars simulated in Band 3, 7 show good results with random residuals and $\chi_{red}^2 \sim 1$. And for Band 5, out of the 18 sources, simulated profiles of 15 pulsars show strong alignment with the observed profiles.

The success of multi-component Gaussian fitting in this analysis supports the theory that pulsar emission is composed of several discrete, potentially overlapping, emission cones or patches \cite{Rankin1983}. Beyond its utility in modeling known interstellar medium (ISM) effects, this simulator is uniquely positioned to address recent fundamental discoveries regarding the pulsar emission mechanism. A striking discovery in the MSP population is that 39$\%$ of these sources exhibit `disjoint' components—emission peaks separated by regions completely lacking in flux—compared to only 3$\%$ of slow pulsars. Recent analysis surmises that these features represent radio emission produced far from the stellar surface, beyond the light cylinder. 

The multi-component Gaussian synthesis engine developed in this work is ideally suited to model this phenomenon. By utilizing the simulation input file configuration, researchers can precisely define these disjoint components as independent Gaussian peaks at wide phase separations. Because the simulator treats each component as a distinct physical entity with its own frequency-dependent scaling laws for height and width, it can recreate the complex Class D (disjoint) profiles now recognized as a hallmark of MSP emission. By generating synthetic epochs with stochastic DM and scattering variations, one can isolate and quantify how these recently discovered LC components contribute to the overall timing residuals of Pulsar Timing Array (PTA) sources. 

Thus, this work will be useful for precise pulsar timing and will be a tool for the Indian Pulsar Timing Array (InPTA) and also aid the researchers in modeling the interstellar medium effects for each pulsar in a synthetic, yet realistic environment. 

%psrchive - freq+ plot after frequency scrunching and dedispersing to give an integrated profile - gives the plot of flux vs phase, frequency vs phase which gives 'luminosity' and the plot of spectral power -  attach results for at least 3 pulsars 
%without dedispersing - shows the effect of time delay due to the DM
%how the spectral power was checked after dedispersion
%effect of scattering and dm smear in the profile at low frequency 
%residue data-model(simulated) at least for 3 to suggest that this works - write how bins made equal and overlapped acc to profile and frequency. Mention that the reduced chi square is near to 1 for almost 15/18 (for which Band 5 data was available), then move on to the band 3 ones.
%residual widths - attach for 3
%how useful for pulsar timing 
%emission mechanism 